\documentclass[12pt,eqsecnum]{article}
\usepackage[dvips]{graphicx}
\usepackage{amssymb}
\usepackage{amsmath}
\usepackage{epstopdf}
\makeatletter

\@addtoreset{equation}{section}
\makeatletter

\newcommand{\D}{{\rm d}}

\newcommand{\dalm}{\kern1pt\vbox{\hrule height 0.9pt\hbox{\vrule width
0.9pt\hskip 2.5pt\vbox{\vskip 5.5pt}\hskip 3pt\vrule width 0.3pt}\hrule height
0.3pt}\kern1pt}

\def\b2hat{ {\hat b}_2 }

\def\be {\begin{equation}}
\def\ee  {\end{equation}}
\def\bea {\begin{eqnarray}}
\def\eea {\end{eqnarray}}

\begin{document}

\begin{titlepage}
\vfill
\begin{flushright}
\today
\end{flushright}

\vfill
%\vskip 1.0cm
\begin{center}
\baselineskip=16pt
{\Large\bf 
Throat quantization of the Schwarzschild-Tangherlini(-AdS) black hole\\
}
\vskip 0.5cm
{\large {\sl }}
\vskip 10.mm
{\bf Gabor Kunstatter${}^{a}$ and Hideki Maeda${}^{b,c}$} \\

\vskip 1cm
{
	${}^a$ Department of Physics, University of Winnipeg and Winnipeg Institute for Theoretical Physics, Winnipeg, Manitoba, Canada R3B 2E9\\
    ${}^b$ Department of Physics, Rikkyo University, Tokyo 171-8501, Japan\\
	${}^c$ Centro de Estudios Cient\'{\i}ficos (CECs), Casilla 1469, Valdivia, Chile. \\
	\texttt{g.kunstatter-at-uwinnipeg.ca, hidekism-at-rikkyo.ac.jp}

     }
\vspace{6pt}
\today
\end{center}
\vskip 0.2in
\par
\begin{center}
{\bf Abstract}
\end{center}
\begin{quote}
Adopting the throat quantization pioneered by Louko and M\"akel\"a, we derive the mass and area spectra for the Schwarzschild-Tangherlini black hole and its anti-de~Sitter (AdS) generalization in arbitrary dimensions.  
We find that the system can be quantized exactly in three special cases: the three-dimensional BTZ black hole, toroidal black holes in any dimension, and five-dimensional Schwarzshild-Tangherlini(-AdS) black holes.  For the remaining cases the spectra are obtained for large mass using the WKB approximation. For asymptotically flat black holes, the area/entropy has an equally spaced spectrum, as expected from previous work.  In the asymptotically AdS case on the other hand, it is the mass spectrum that is equally spaced.
Our exact results for the BTZ black hole mass with Dirichlet and Neumann boundary conditions are consistent with the spacing in the spectra of the corresponding  operators in the dual CFT.
\vfill
% \hrule width 5.cm
\vskip 2.mm
\end{quote}
\end{titlepage}

%<<<<<<<<<<<<< PACS NUMBER >>>>>>>>>>>>>>>%
%\pacs{
%04.50.-h 	Higher-dimensional gravity and other theories of gravity
%04.50.Gh 	Higher-dimensional black holes, black strings, and related objects 
%04.60.-m 	Quantum gravity
%04.60.Ds 	Canonical quantization 
%04.60.Kz 	Lower dimensional models; minisuperspace models 
%} 

% RUP-13-15, CECS-PHY-13/09

%\maketitle
%\section{}
%\subsection{}

\tableofcontents

\newpage

%======================================%
%<<<<<<<<<<<< SECTION I  >>>>>>>>>>>>>>%
%======================================%
\section{Introduction}
Well known theorems~\cite{singularities} imply that singularities must generically appear somewhere in a universe governed by the classical theory of general relativity.
It is generally believed that the classical theory breaks down and quantum aspects of gravity dominate around such singularities, thereby resolving them.
Thus a complete description of the universe requires a quantum theory of gravity.

One of the oldest approaches to finding such a quantum theory is canonical quantum gravity which is based on the  ADM decomposition~\cite{adm} of the spacetime. The ADM decomposition reveals that general relativity is a constrained dynamical system, with two types of constraints: a set of momentum constraint $H^i=0$, which generate spatial diffeomorphisms and the Hamiltonian constraint $H=0$, which generates time reparametrizations. While the momentum constraints are linear in the momenta of the fields, the Hamiltonian constraint is highly nonlinear and difficult to solve. This makes a completely reduced quantization of the full system, in which a coordinate system is chosen and the constraints solved prior to quantization, highly problematic. 

The most common approach is therefore Dirac's method of quantization for constrained systems~\cite{Dirac}. In Dirac quantization, the constraints become operators that annihilate physical states which are given in the Schr\"odinger approach by functionals $\Psi[h_{ij}]$  of the spatial metric $h_{ij}$. The basic equations in canonical quantum gravity are then the momentum constraints ${\hat H}^i\Psi=0$ and the Hamiltonian constraint ${\hat H}\Psi=0$.
The momentum constraints can be formally solved (and in some cases explicitly solved) simply by ensuring that the wave functional $\Psi$ is a diffeomorphism invariant functional of $\tilde{h}_{ij}$, i.e. its argument takes values in the quotient space of $h_{ij}$ under the action of the group of spatial diffeomorphisms. The remaining Hamiltonian constraint, the so-called the Wheeler-de~Witt equation,
is very difficult to solve in full generality. Loop quantum gravity is one program that attempts to do this via a suitable choice of variables~\cite{lqg}.

An alternative way to make the problem tractable is to consider a restricted class of spacetimes by imposing a symmetry.
Even in such a midisuperspace approach, the system still generally describes an infinite number of degrees of freedom and remains exceedingly difficult to quantize.

There is, however, one exception, namely spherical symmetry. The Birkhoff's theorem in general relativity guarantees that when spherical symmetry is imposed, one is left with only a single pair of physical gravitational phase space degrees of freedom.
In a seminal analysis, Kucha\v{r}~\cite{kuchar94} showed for spherically symmetric black-hole spacetimes that this pair can be taken to be the ADM mass of the black hole and its conjugate, the Schwarzschild time separation at spatial infinity. More specifically, Kucha\v{r} was able to construct explicitly  under appropriate fall-off conditions the canonical transformation from general ADM variables to the geometrical variables consisting of the areal radius $r(x,t)$, the Misner-Sharp mass function $M(x,t)$~\cite{ms1964} and their canonical conjugates. 
In terms of this parametrization the constraints are easily solved. On the constraint surface, the Misner-Sharp mass depends only on the time coordinate $t$, namely $M(x,t)={\bf m}(t)$. Moreover the conjugate to the areal radius vanishes, so that $r$ can be specified arbitrarily in terms of the spatial coordinates: such a choice merely fixes the diffeomorphism invariance. 
On the constraint surface one therefore obtains a fully reduced action that depends only ${\bf m}(t)$ and its conjugate ${\bf p}$. The reduced action takes the form:
\be
I[{\bf m},{\bf p}]=\int dt \biggl[{\bf p}(t) \dot{{\bf m}}(t) - (N_+-N_-){\bf m}(t)\biggl], \label{intro}
\ee
where  { the prescribed functions} $N_\pm(t)$ are the values of the lapse at either ends of the spatial slice and { are not varied}.
The Hamilton equation for ${\bf m}$ is then simply ${\dot {\bf m}}=0$, whose solution requires ${\bf m}$ to be a constant equal to the ADM mass.
By Birkhoff's theorem,  ${\bf m}$ carries all the information about the (local) geometry of the classical solutions.
%Canonical quantization of this system shows that mass is actually an eigenvalue for the wave function of the spacetime.

While this analysis provides an elegant illustration of the relationship between the geometrical and physical content of the classical theory, it is not particularly useful for quantization. In essence, the action for any classical Hamiltonian system with only two phase space degrees of freedom can formally be put into the above form via a suitable canonical transformation, the energy $E$ playing the role of ${\bf m}$. Unless one knows something further about the space of solutions (e.g. periodicity), one has very few clues as to how to proceed in order to obtain a spectrum for $E$, or ${\bf m}$ as in the present case.  What is needed is a physical set of phase space variables for which the boundary conditions are known. Construction of a self-adjoint Hamiltonian in terms of these variables and boundary conditions then yields the energy spectrum. 

This is precisely what  Louko and M\"akel\"a~\cite{LM96} achieved in a very elegant and clear analysis by introducing the throat quantization method for the Schwarzschild black hole.
They chose as the physical Lagrangian degree of freedom the minimum radius of the throat of the Einstein-Rosen bridge associated with an asymptotically flat eternal black hole and expressed the dynamics in terms of the proper time of a comoving observer.
The corresponding geodesic represents the dynamics of the wormhole throat in the Schwarzschild spacetime foliated by spacelike hypersurfaces and its orbit is from the white-hole singularity to the black-hole singularity through the bifurcation two-sphere.
They explicitly constructed the Hamiltonian for the throat dynamics and showed how to obtain it by a time dependent canonical transformation from Kucha\v{r}'s action (\ref{intro}). 
This is a time-independent canonical transformation and the Hamiltonian, which is the ADM mass in this approach, is a constant of motion in the throat dynamics.
Using the WKB approximation, they derived the mass spectrum of the Schwarzschild black hole in the large mass region.
Their result was that the area and hence the Bekenstein-Hawking entropy are equally spaced in such a regime, in keeping with the general predictions of Bekenstein and Mukhanov based on heuristic arguments~\cite{bekenstein}. The spacing was however different from that of~\cite{bekenstein}.

We note that while many other black hole quantization schemes exist in the literature \cite{Kastrup1996,Barvinsky1997,Vaz1999}, the method of \cite{LM96} is unique in that it provides a rigorous quantization of geometrical variables in terms of a physical time parameter using a Hamiltonian that is obtained via a canonical transformation from Kucha\v{r}'s ``true'' reduced Hamiltonian~\cite{kuchar94}.

The purpose of the present paper is to generalize the result of \cite{LM96} to arbitrary dimensions with or without a negative cosmological constant.
We also apply the method to topological black holes. 
We show that the energy/mass spectra are in general bounded below. 
Remarkably, the resulting quantum system can be exactly quantized in a few special cases: the three-dimensional BTZ black hole, the five-dimensional black hole, and the toroidal black hole. 
For the remaining cases the WKB approximation can be rigorously applied to yield the semi-classical spectrum. 
The general result is that, for asymptotically flat black holes, the area spectrum is equally spaced,  while for asymptotically AdS black holes it is the mass spectrum that is equally spaced. 
The asymptotically flat case agrees semi-classically with the spectra derived for higher-dimensional black holes using different methods (polymer quantization) in \cite{peltola}. 
One significant result is that the exact spectrum we obtain for the BTZ black hole using Dirichlet and Neumann boundary conditions agrees precisely with the spacing of the spectrum obtained previously by Birmingham and Carlip~\cite{carlip} from the quasinormal mode spectrum of the BTZ black hole. 
Their result, was consistent with the spectrum obtained from microscopic D-brane physics~\cite{stringy}. 
It is encouraging that the simple model considered here yields a spectrum that is consistent with the prediction from the microscopic theory in the one case where such a microscopic description exists.

The outline of the present paper is as follows.
In section~\ref{sec:throat}, the throat quantization method is explained.
In section~\ref{sec:exact}, we derive the mass and entropy spectra in the case where the Schr\"odinger equation is solved analytically.
Section~\ref{sec:WKB} is devoted to deriving the spectra for large mass via the WKB approximation.
Our conclusions and discussions are summarized in section~\ref{sec:summary}.
The operator-ordering that we adopt in the present paper is explained in appendix A, the general proof that the energy is bounded below is given in Appendix B, while the quantization of the simple harmonic oscillator on the half line is given in Appendix C.
Our basic notation follows~\cite{wald}.
The convention for the Riemann curvature tensor is $[\nabla _\rho ,\nabla_\sigma]V^\mu ={R^\mu }_{\nu\rho\sigma}V^\nu$ and $R_{\mu \nu }={R^\rho }_{\mu \rho \nu }$.
The Minkowski metric is taken as diag$(-,+,\cdots,+)$, and Greek indices run over all spacetime indices.
We adopt the units in which only the $n$-dimensional gravitational constant $G_n$ is retained.

%======================================%
%<<<<<<<<<<<< SECTION I  >>>>>>>>>>>>>>%
%======================================%
\section{Throat quantization of symmetric black holes}
\label{sec:throat}

\subsection{Preliminaries}
We consider general relativity with a cosmological constant $\Lambda$ in arbitrary $n(\ge 3)$ dimensions, whose action is given by 
\begin{align}
I=&\frac{1}{2\kappa_n^2}\int \D ^nx\sqrt{-g}(R-2\Lambda)+I_{\partial{\cal M}},\label{action}
\end{align}
where $\kappa_n := \sqrt{8\pi G_n}$ and $I_{\partial{\cal M}}$ is the York-Gibbons-Hawking boundary term.
Varying the above action, we obtain the Einstein field equations:
\begin{align} 
R_{\mu\nu}-\frac12 g_{\mu\nu}R+\Lambda g_{\mu\nu}=0. \label{beqL}
\end{align} 
Assume an $n$-dimensional warped product spacetime $({\cal M}^n,g_{\mu\nu}) \approx ({M}^2,g_{AB})\times ({K}^{n-2},\gamma_{ab})$ with the general metric 
\begin{eqnarray}
g_{\mu\nu}(x)dx^\mu dx^\nu=g_{AB}({\bar y})d{\bar y}^A d{\bar y}^B+r({\bar y})^2\gamma_{ab}(z)dz^adz^b,
\label{eq:structure}
\end{eqnarray}
where indices run as $A,B=0,1$ and $a,b=2,3,\cdots,n-1$.
$g_{AB}$ is an arbitrary Lorentz metric on $({M}^2,g_{AB})$ and $R({\bar y})$ is a scalar function on $({M}^2,g_{AB})$.
$\gamma_{ab}$ is the unit metric on the $(n-2)$-dimensional maximally symmetric space $({K}^{n-2},\gamma_{ab})$ with sectional curvature $k=1,0,-1$. 
{ Note that in three dimensions ($n=3$) the internal space is one dimensional and necessarily flat ($k=0$) .} 
The generalized Misner-Sharp mass~\cite{ms1964,mn2008} in this system is defined by
\begin{align}
M:=& \frac{(n-2)V_{n-2}^{(k)}}{2\kappa_n^2}r^{n-3}\biggl(\frac{r^2}{l^2}+(k-(Dr)^2)\biggl),\label{qlm}\\
l^2:=&-\frac{(n-1)(n-2)}{2\Lambda}, \label{lambdatil}
\end{align}  
where $(Dr)^2:=(D_A r)(D^A r)$.
The constant $V_{n-2}^{(k)}$ represents the volume of $({K}^{n-2},\gamma_{ab})$ if it is compact  and otherwise arbitrary positive.
$M$ reduces to the ADM (Arnowitt-Deser-Misner) mass at spacelike infinity in the asymptotically flat spacetime~\cite{hayward1996,mn2008}.

\subsection{Vacuum black holes}
\begin{figure}[htbp]
\begin{center}
%\rotatebox{-90}{
\includegraphics[width=0.8\linewidth]{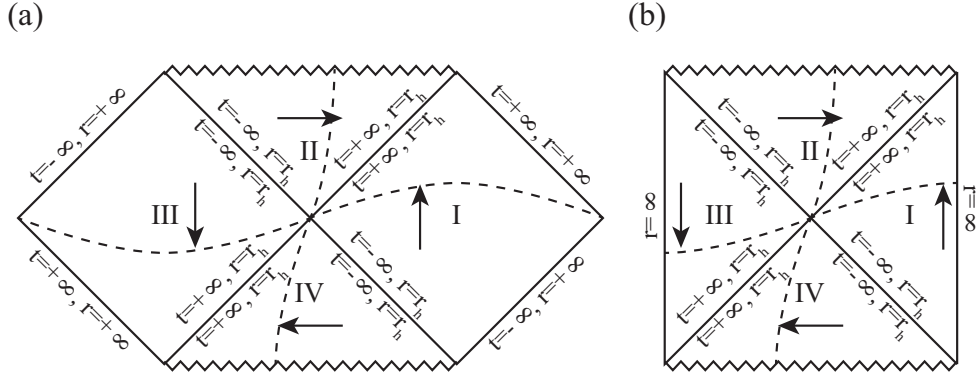}
%\subfigure[]{\includegraphics[width=0.7\linewidth]{Roberts-lambda1.eps}}
%\subfigure[]{\includegraphics[width=0.7\linewidth]{Roberts-lambda2.eps}}
%}
\caption{\label{SingleBH} Penrose diagrams for the Schwarzschild-Tangherlini-type black hole (\ref{f-vacuum}) with a single horizon with (a) $\Lambda=0$ and (b) $\Lambda<0$. A dashed curve in each portion of the spacetime represents a constant $t$ hypersurface. The arrows show the directions of increasing   $t$.}
\end{center}
\end{figure}

\begin{figure}[htbp]
\begin{center}
%\rotatebox{-90}{
\includegraphics[width=0.5\linewidth]{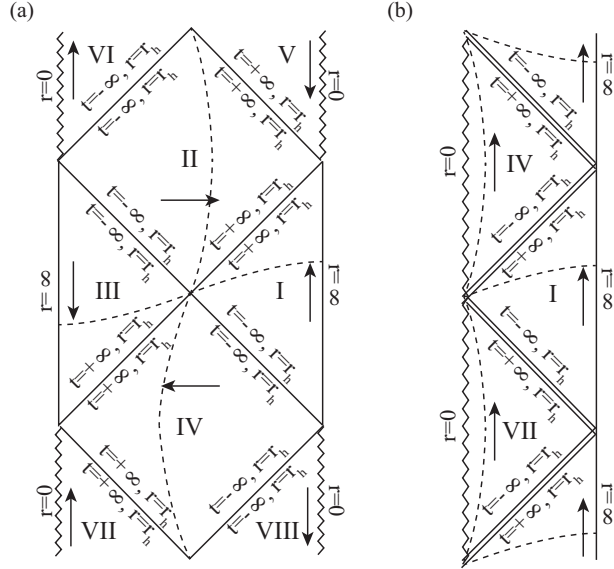}
%\subfigure[]{\includegraphics[width=0.7\linewidth]{Roberts-lambda1.eps}}
%\subfigure[]{\includegraphics[width=0.7\linewidth]{Roberts-lambda2.eps}}
%}
\caption{\label{SingleBH2} Penrose diagrams for the Schwarzschild-Tangherlini-type black hole (\ref{f-vacuum}) with $k=-1$ and $\Lambda<0$ with (a) two non-degenerate horizons and (b) one degenerate horizon. A dashed curve in each portion of the spacetime represents a constant $t$ hypersurface. The arrows show the increasing directions of $t$.}
\end{center}
\end{figure}

In the vacuum case, $M$ is constant.  
If $r$ is not constant, the most general solution is the following Schwarzschild-Tangherlini-type solution\footnote{If $r$ is constant, the Nariai (anti-Nariai) solution is possible for $k=1$ ($k=-1$) and positive (negative) $\Lambda$, which is a cross product of a two-dimensional de~Sitter (anti-de~Sitter) spacetime and a $(n-2)$-dimensional space of positive (negative) constant curvature~\cite{nariai,jiri}.}:
\begin{align}
ds^2=&-f(r)dt^2+f(r)^{-1}dr^2+r^2\gamma_{ab}dz^a dz^b, \label{f-vacuum} \\
f(r) =&k-\frac{2\kappa_n^2M}{(n-2)V_{n-2}^{(k)}r^{n-3}}+\frac{r^2}{l^2}.
\end{align}
The range of parameter $M$ that gives black-hole configurations is summarized in Table~\ref{table:horizon} and the possible Penrose diagrams for the black holes are shown in Figs~\ref{SingleBH} and \ref{SingleBH2}~\cite{tm2005}.
%-------------- TABLE ---------------%
\begin{table}[h]
\caption{\label{table:horizon} Existence of the outer Killing horizon in the Schwarzschild-Tangherlini-type spacetime (\ref{f-vacuum}) for $n\ge 4$. Here $M_{\rm ex}:=(n-2)V_{n-2}^{(k)}kr_{\rm ex}^{n-3}/[(n-1)\kappa_n^2]$ is the mass for the extremal horizon, where $r_{\rm ex}:=\sqrt{(n-2)(n-3)k/2\Lambda}$ is its horizon radius.
For $n=3$ (and $k=0$), an outer Killing horizon exists for $M>0$ and $\Lambda<0$.
}
\begin{center}
\begin{tabular}{l@{\qquad}c@{\qquad}c@{\qquad}c}
\hline \hline
  & $k=1$ & $k=0$ & $k=-1$   \\\hline
$\Lambda=0$ & $M>0$ & Not applicable & Not applicable \\ \hline
$\Lambda>0$ & $0<M\le M_{\rm ex}$ & Not applicable & Not applicable \\ \hline
$\Lambda<0$ & $M>0$ & $M>0$ & $M\ge M_{\rm ex}(<0)$ \\
\hline \hline
\end{tabular}
\end{center}
\end{table} 
%------------------------------------%
%------------<fig>---------------------------
%--------------<fig>-----------------------
%------------<fig>---------------------------
%--------------<fig>-----------------------

The relation between the mass parameter $M$ and the horizon radius $r_{\rm h}$ is 
\begin{align}
M=\frac{(n-2)V_{n-2}^{(k)}r_{\rm h}^{n-3}}{2\kappa_n^2}\biggl(k+\frac{r_{\rm h}^2}{l^2}\biggl).
\end{align}
The Wald entropy $S$ of a black hole~\cite{wald1993} is given by 
\begin{align}
S=&\frac{2\pi}{\kappa_n^2}V_{n-2}^{(k)}r_{\rm h}^{n-2}.
\end{align}
For $k=1$ and $\Lambda=0$, the resulting relation between the mass and the entropy is 
\begin{align}
S=\frac{2\pi}{\kappa_n^2}V_{n-2}^{(1)}\biggl(\frac{2\kappa_n^2M}{(n-2)V_{n-2}^{(1)}}\biggl)^{(n-2)/(n-3)}, \label{entropy}
\end{align}
while for $k=0$ and $\Lambda<0$, it is 
\begin{align}
S=&\frac{2\pi}{\kappa_n^2}V_{n-2}^{(0)}\biggl(\frac{2\kappa_n^2l^2M}{(n-2)V_{n-2}^{(0)}}\biggl)^{(n-2)/(n-1)}.
\end{align}

\subsection{Reduced action in geometrodynamics}

The geometrodynamics of this vacuum can be obtained by considering the midisuperspace containing the 2-metric in ADM coordinates $(t,x)$ on $(M^2,g_{AB})$:
\begin{eqnarray}
ds_{(2)}^2=g_{AB}d{\bar y}^Ad{\bar y}^B=-N(t,x)^2dt^2+{\bar \Lambda}(t,x)^2(dx+N_x(t,x)dt)^2.\label{ADM}
\end{eqnarray}
and the areal radius $r(t,x)$.
The standard Hamiltonian analysis reveals the lapse $N$ and shift $N_x$ to be Lagrange multipliers enforcing a Hamiltonian and diffeomorphism constraint, respectively.
Kucha\v{r}~\cite{kuchar94} explicitly constructed the canonical transformation, given asymptotically flat boundary conditions in four dimensions, from the standard ADM variables, to the areal radius and the Misner-Sharp mass $M$ as phase space variables.
This has been generalized to arbitrary dimensions an general $k$ in \cite{dl2009}.
In terms of $M$, the Hamiltonian constraint is simply $\partial M/\partial x=0$, whose solution is $M={\bf m}(t)$.
Putting this solution back to the Lagrangian, one obtains the reduced action~\cite{kuchar94,dl2009,kmt2013}:
\begin{align}
I[{\bf m},{\bf p}]=&\int \D t\biggl({\bf p}{\dot {\bf m}}-(N_+ - N_-){\bf m}\biggl), \label{reduced-action}
\end{align}
which has only a finite number (two) of phase space degrees of freedom.
Here $N_+(t)$ and $N_-(t)$, the Lapse functions at right and left infinities, respectively are fixed and hence not varied.
${\bf p}(t)$ is its momentum conjugate of ${\bf m}$.
${\bf p}$ represents the difference of the asymptotic Killing times between the right and left infinities on the constant $t$ hypersurface because the Hamilton equation for ${\bf p}$ is ${\dot {\bf p}}=-N_++N_-$.
The Hamilton equation for ${\bf m}$ is then ${\dot {\bf m}}=0$, whose solution is ${\bf m}$ is constant, the ADM mass. By Birkhoff's theorem,  ${\bf m}$ carries all the information about the (local) geometry of the classical solutions. Kucha\v{r} also showed that the conjugate to ${\bf m}$ is the Schwarzschild time separation at the two ends of the spatial (constant time) slice. 

As stated in the Introduction, the above form of the reduced action is not particularly well suited for quantization. One can in fact put the action for any one-dimensional Lagrangian system into this form. The only information it contains is that the energy ${\bf m}$ is a constant of the motion. More information is needed about the solution space in order to get a non-trivial quantum theory. This requires transforming to a different set of physical observables with non-trivial dynamics and known ranges. 

Following the work of Louko and M\"akel\"a~\cite{LM96}, we choose as the relevant physical observable the minimum radius $a$ along any spatial slice of the throat of the Einstein-Rosen bridge. The dynamics of the wormhole throat radius, when expressed as a function of the proper time of a comoving observer provides a coordinate invariant basis for the quantization of the black-hole geometry under consideration. To connect the wormhole dynamics to the ADM analysis above,
we first set $N_+=1$ and $N_-=0$, which means that $t$ coincides with the one in the right-hand asymptotic Minkowski region, up to the additive constant, while the hypersurface at the left-hand asymptotic infinity is frozen for all $t$. Moreover, we adjust the slicing so that $t$ coincides up to an additive constant with the proper time of a comoving observer sitting at the minimum radius. Note that this comoving observer is within the black hole interior and ``comoving'' implies that the Schwarzschild time remains fixed.
The reduced action becomes $I=\int \D t({\bf p}{\dot {\bf m}}-{\bf m})$, for which the Hamiltonian is simply ${\bf m}$, which harmonizes with the fact that ${\bf m}$ is the ADM mass as the conserved energy in the right-hand side asymptotic Minkowski time evolution.
 
 In the next subsections, we will show that there exists in general a canonical transformation from the geometrodynamical variables to the wormhole throat variables $a=a({\bf m},{\bf p})$ and $p_a=p_a({\bf m},{\bf p})$, where $p_a$ is the momentum conjugate of $a$.
The canonical transformation $({\bf m},{\bf p})\to (a,p_a)$ is time-independent and therefore the Hamiltonian $H(a,p_a)$ for the new set of variable $(a,p_a)$ is simply ${\bf m}(a,p_a)$.

\subsection{Throat dynamics}
In order to perform the throat quantization, we will make another canonical transformation from the Kucha\v{r} action (\ref{reduced-action}) to another one.
The terminology "throat quantization" comes from the physical interpretation of the dynamics represented by the new canonical variables.
Actually, it represents the dynamics of the wormhole throat in the maximally extended Schwarzschild-Tangherlini-type black-hole spacetime, which will be explained below.

The metric in the Schwarzschild-Tangherlini-type spacetime is given by 
\begin{align}
ds^2=-f({\bf m},r)dT^2+f({\bf m},r)^{-1}dr^2+r^2\gamma_{ij}dz^idz^j,
\end{align}
where ${\bf m}$ is the Misner-Sharp mass and  the metric function $f({\bf m},r)$ is determined algebraically by  
\begin{align}
&{\bf m} = \frac{(n-2)V_{n-2}^{(k)}}{2\kappa_n^2}r^{n-3}\biggl(\frac{r^2}{l^2}+k-f({\bf m},r)\biggl), \nonumber \\
\to &f({\bf m},r) =k- \frac{2\kappa_n^2{\bf m}}{(n-2)V_{n-2}^{(k)}r^{n-3}}+\frac{r^2}{l^2}.
\label{eq:m}
\end{align}  

Consider radial timelike geodesics $T=T(\tau), r=a(\tau)$ inside the Killing horizon where $f({\bf m},r)<0$ is satisfied, along which
\begin{align}
-1=-f({\bf m},a(\tau))\biggl(\frac{dT}{d\tau}\biggl)^2+f({\bf m},a(\tau))^{-1}\biggl(\frac{da}{d\tau}\biggl)^2
\end{align}
holds, where $\tau$ is a proper time along the geodesic.
We now focus on the particular subset of this family of geodesics corresponding to the path of a comoving observer sitting stationary ($T=$constant) at the narrowest point of the wormhole throat. (Recall that in the black hole interior $T$ is a spacelike coordinate.) 
Such a geodesic equation is given by 
\begin{align}
\frac{d^2a}{d\tau^2}=-\frac{(n-3)\kappa_n^2{\bf m}}{(n-2)V_{n-2}^{(k)}a^{n-2}}-\frac{a}{l^2}, \label{check1}
\end{align}
The radial coordinate $r=a(\tau)$, which is by the above the minimum radius of the wormhole, then has a velocity with respect to the proper time of such an observer as follows:
\begin{align}
v:=&\frac{da}{d\tau}=\pm \sqrt{-f({\bf m},a)}, \label{def-v}
\end{align}
where the plus sign and minus signs are taken in the white-hole (past trapped) and black-hole (future trapped) regions, respectively.
Integrating the above equation, we obtain
\begin{align}
\tau-\tau_0=\pm \int^a_0\frac{d{\bar a}}{\sqrt{-f({\bf m},{\bar a})}},
\end{align}
where constant $\tau_0$ is the proper time at $a(\tau_0)=0$.
Without loss of generality we can set $\tau_0=0$ by reparametrization of $\tau$ and then we have
\begin{align}
\tau(a)=\pm \int^a_0\frac{d{\bar a}}{\sqrt{-f({\bf m},{\bar a})}}, \label{tau-def}
\end{align}
Then, $\tau>(<)0$ is satisfied for the plus (minus) sign.
For a spacetime characterised by a particular value of ${\bf m}$, the range of $a$ is $0\le a\le a_{\rm h}$, where $a_{\rm h}$ is the horizon radius defined by $f({\bf m}, a_{\rm h})=0$, and corresponds to a turning point in the solution. The corresponding range of $\tau$ is $0\le\tau\le \tau(a_{\rm h})$ for the plus sign and $-\tau(a_{\rm h})\le\tau\le 0$ for the minus sign.
The total proper time required for the throat to grow from zero radius at the white hole singularity through its maximum value $a_{\rm h}$ and then recollapse to zero radius is: 
\begin{align}
\tau_{\rm tot}({\bf m})= 2\int^{a_{\rm h}}_0 \frac{d{\bar a}}{\sqrt{-f({\bf m},{\bar a})}}, \label{tau-def}
\end{align}
For a solar mass Schwarzschild mass black hole, this gives the textbook value of the order of microseconds. This is the motion we wish to quantize. The first step is to express the  total energy ${\bf m}$ in terms of the throat variables, which we do in the next subsection.

\subsection{Canonical transformation}
Here we make a canonical transformation from the Kucha\v{r} variables $({\bf m},{\bf p})$ to another set of variables $(a,p_a)$, where $p_a$ is the momentum conjugate to $a$.
Inserting the expression for the velocity of the throat (\ref{def-v}) into Eq.~(\ref{eq:m}), which in turn is the Hamiltonian for the geometrodynamic system, we get: 
\begin{align}
{\bf m} =& \frac{(n-2)V_{n-2}^{(k)}}{2\kappa_n^2}a^{n-3}\biggl(\frac{a^2}{l^2}+k-f({\bf m},a)\biggl)=:H(a,v)
\end{align} 
Given the above expression for $H(a,v)={\bf m}$ in terms of $a$ and its velocity $v$, it is straightforward to construct a canonical transformation that relates $({\bf m}, {\bf p})$ to $(a,p_a)$.

By definition of $v$ (\ref{def-v}), the Hamiltonian $H(a,p_a)$ with new variables satisfies the Hamilton equation
\begin{align}
v=&\frac{\partial H(a,p_a)}{\partial p_a}\\
 =&\frac{\partial {\bf m}(a,p_a)}{\partial p_a}. 
 \label{v-H}
\end{align}
For this purpose, we consider a generating function $G(a,{\bf m})$ for this transformation, satisfying
\begin{align}
{\bf p}={\bf p}(a,{\bf m})=-\frac{\partial G}{\partial{\bf m}}\biggl|_{a}, \quad p_a=p_a(a,{\bf m})=\frac{\partial G}{\partial a}\biggl|_{{\bf m}}.
\end{align}
The integrability condition ensuring the existence of $G(a,{\bf m})$ is 
\begin{align}
\frac{\partial {\bf p}(a,{\bf m})}{\partial a}\biggl|_{{\bf m}}=-\frac{\partial p_a(a,{\bf m})}{\partial{\bf m}}\biggl|_{a}.
\end{align}
Hence we define $p_a=p_a({\bf m},{\bf p})$ such that
\begin{align}
{\bf p}(a,{\bf m})=\int_a^{a_{\rm h}({\bf m})}d{\bar a}\frac{\partial p_a({\bar a},{\bf m})}{\partial{\bf m}}\biggl|_{\bar a}, \label{def-p}
\end{align}
where $a_{\rm h}({\bf m})$ is the horizon radius determined by $f({\bf m},a_{\rm h})=0$.

Now ${\bf p}={\bf p}(a,{\bf m}),p_a=p_a(a,{\bf m})$ means ${\bf m}={\bf m}(a,p_a),{\bf p}={\bf p}(a,p_a)$ and hence 
\begin{align}
\frac{\partial p_a(a,{\bf m})}{\partial{\bf m}}\biggl|_{a}=\biggl(\frac{\partial {\bf m}(a,p_a)}{\partial p_a}\biggl|_{a}\biggl)^{-1}
\end{align}
is satisfied.
Using this and $H(a,p_a):={\bf m}(a,p_a)$, we obtain
\begin{align}
\frac{\partial p_a(a,{\bf m})}{\partial{\bf m}}\biggl|_{a}=\biggl(\frac{\partial H(a,p_a)}{\partial p_a}\biggl|_{a}\biggl)^{-1}=\frac{1}{v(a,p_a)}=\pm\frac{1}{\sqrt{-f({\bf m},a)}},
\end{align}
where we used Eq.~(\ref{v-H}).
Finally, Eq.~(\ref{def-p}) reduces to the following relation:
\begin{align}
|{\bf p}|=&\int_a^{a_{\rm h}({\bf m})}\frac{db}{\sqrt{-f({\bf m},b)}}.
\end{align}
For a given classical solution, the range of $a$ is $0\le a\le a_{\rm h}$ while the range of ${\bf p}$ is determined by the above equation.
Comparing the above expression with Eq.~(\ref{tau-def}), we find that $-\tau(a_{\rm h})\le{\bf p}\le \tau(a_{\rm h})$.

In order to define the transformation completely, we need another relation between $({\bf m},{\bf p})$ and $(a,p_a)$. From (\ref{v-H}) we have:
\bea
p_a &=& \int \frac{d{\bf m}}{v}\nonumber\\
 &=& \int \frac{d{\bf m}}{\sqrt{-f({\bf m},a)}}\nonumber\\
  &=& \mbox{sgn}({\bf p})\sqrt{-\biggl(\frac{(n-2)V_{n-2}^{(k)}}{\kappa_n^2}\biggl)^2a^{2(n-3)}f({\bf m},a)}, \label{def-pa}
\eea
This finally allows us to express the Hamiltonian/mass in terms of the canonical variables $a,p_a$:
\be
{\bf m}(a,p_a)=\frac{(n-2)V_{n-2}^{(k)}a^{n-3}}{2\kappa_n^2}\biggl[\biggl(\frac{(n-2)V_{n-2}^{(k)}}{\kappa_n^2}\biggl)^{-2}p_a^2a^{-2(n-3)}+k+\frac{a^2}{l^2}\biggl].
\ee
We confirm the relation (\ref{v-H}) holds:
\begin{align}
\frac{\partial H}{\partial p_a}=\frac{\partial {\bf m}}{\partial p_a}=&\frac{(n-2)V_{n-2}^{(k)}a^{n-3}}{2\kappa_n^2}\biggl(\frac{(n-2)V_{n-2}^{(k)}}{\kappa_n^2}\biggl)^{-2}2p_aa^{-2(n-3)} \nonumber \\
=&\mbox{sgn}({\bf p})\sqrt{-f({\bf m},a)}={\dot a}.
\end{align}

In summary, $({\bf m},{\bf p})\to (a,p_a)$ is a canonical transformation, where
\begin{align}
|{\bf p}|=&\int_a^{a_{\rm h}({\bf m})}\frac{db}{\sqrt{-f({\bf m},b)}},\\
p_a=&\mbox{sgn}({\bf p})\sqrt{-\biggl(\frac{(n-2)V_{n-2}^{(k)}}{\kappa_n^2}\biggl)^2a^{2(n-3)}f({\bf m},a)},\\
f({\bf m},a) :=&k- \frac{2\kappa_n^2{\bf m}}{(n-2)V_{n-2}^{(k)}a^{n-3}}+\frac{a^2}{l^2}
\end{align}
and the new Hamiltonian $H(a,p_a)$ is 
\begin{align}
H=\frac{(n-2)V_{n-2}^{(k)}a^{n-3}}{2\kappa_n^2}\biggl[\biggl(\frac{(n-2)V_{n-2}^{(k)}}{\kappa_n^2}\biggl)^{-2}p_a^2a^{-2(n-3)}+k+\frac{a^2}{l^2}\biggl]. \label{hamiltonian}
\end{align}
The new action for the throat dynamics is 
\begin{align}
I=&\int \D t\biggl(p_a{\dot a}-H(a,p_a)\biggl). \label{newaction}
\end{align}
The turning point $a=a_{\rm h}$ is the maximum value of $a$ for a given value of ${\bf m}$. Since the domain of ${\bf m}$ is ${\bf m}>{\bf m}_{\rm crit}$, the corresponding domain of $a$ is $0\le a <\infty$. The domain of its conjugate $p_a$ is $-\infty<p_a<\infty$.

One can confirm that the Hamiltonian (\ref{hamiltonian}) certainly provides the geodesic equation for the wormhole throat (\ref{check1}), where $t$ is identified as the proper time $\tau$ of the wormhole throat.
The Hamilton's equations ${\dot a}=\partial H/\partial p_a$ gives
\begin{align}
p_a=\frac{(n-2)V_{n-2}^{(k)}}{\kappa_n^2}a^{n-3}{\dot a}.
\end{align}
Using this and Eq,~(\ref{hamiltonian}), we obtain
\begin{align}
{\dot a}^2=\frac{2\kappa_n^2}{(n-2)V_{n-2}^{(k)}a^{n-3}}H(a,p_a)-k-\frac{a^2}{l^2}.
\end{align}
Finally the other Hamilton's equation ${\dot p}_a=-\partial H/\partial a$ provides Eq.~(\ref{check1}) as
\begin{align}
{\ddot a}=&-\frac12\biggl((n-3)a^{-1}{\dot a}^2+(n-3)ka^{-1}+\frac{(n-1)a}{l^2}\biggl) \nonumber \\
=&-\frac{(n-3)\kappa_n^2H}{(n-2)V_{n-2}^{(k)}a^{n-2}}-\frac{a}{l^2}.
\end{align}

\subsection{Hamiltonian operator}
In throat quantization, we quantize the system (\ref{newaction}).
Replacing $p_a$ in the Hamiltonian (\ref{hamiltonian}) by ${\hat p}_a=-i\hbar \partial/\partial a$, we obtain the Schr\"odinger equation for the wave function $\Psi=\Psi(t,a)$:
\begin{align}
{\hat H}\Psi =i\hbar\frac{\partial\Psi}{\partial t}. \label{Schro}
\end{align}  
Putting $\Psi(t,a)=e^{-iEt/\hbar}\psi(a)$, where $E$ is a constant, we obtain 
\begin{align}
{\hat H}\psi =E\psi.
\label{eq:se 1}
\end{align}  
Here $E$ represents the mass of the black hole because the Hamiltonian of the system (\ref{newaction}) is the ADM mass in the throat dynamics.

For reasons explained in Appendix A, we adopt the natural factoring ordering:
\begin{align}
{\hat H} =\frac{(n-2)V_{n-2}^{(k)}}{2\kappa_n^2}\biggl\{\frac{a^{n-1}}{l^2}+ka^{n-3}-\biggl(\frac{\kappa_n^2}{(n-2)V_{n-2}^{(k)}}\biggl)^2\frac{\hbar^2}{a^{(n-3)/2}}\frac{d}{da}\biggl(\frac{1}{a^{(n-3)/2}}\frac{d}{da}\biggl)\biggl\}.
\end{align}  
which is Hermitian with respect to the measure  $\mu(a)=a^{(n-3)/2}$ for the inner product of the wave functions $\Psi$ and $\Phi$ as defined by
\begin{align}
\langle \Psi|\Phi\rangle:=\int^\infty_0\Psi^*\Phi\mu(a)\D a.
\end{align}  
Louko and M\"akel\"a considered the more general measure $\mu(a) = a^\beta$~\cite{LM96}, with corresponding Hermitian ordering of the momentum term in the Hamiltonian.

Defining $x:=a^{(n-1)/2}$, we obtain
\begin{align}
{\hat H} =\frac{(n-2)V_{n-2}^{(k)}}{2\kappa_n^2}\biggl[\frac{x^2}{l^2}+kx^{2(n-3)/(n-1)}-\hbar^2\biggl(\frac{(n-1)\kappa_n^2}{2(n-2)V_{n-2}^{(k)}}\biggl)^2\frac{d^2}{dx^2}\biggl].
\end{align}  
The Schr\"odinger equation ${\hat H}\psi=E\psi$ is then written as
\begin{align}
\frac{(n-2)V_{n-2}^{(k)}}{2\kappa_n^2}\biggl[\frac{x^2}{l^2}+kx^{2(n-3)/(n-1)}-\hbar^2\biggl(\frac{(n-1)\kappa_n^2}{2(n-2)V_{n-2}^{(k)}}\biggl)^2\frac{d^2}{dx^2}\biggl]\psi=E\psi. \label{sch1}
\end{align}  
Comparing with 
\begin{align}
\biggl(-\frac{\hbar^2}{2m}\frac{d^2}{dx^2}+V(x)\biggl)\psi=E\psi,
\end{align}  
we identify the effective mass and potential as
\begin{align}
V(x)\equiv &\frac{(n-2)V_{n-2}^{(k)}}{2\kappa_n^2}\biggl(\frac{x^2}{l^2}+kx^{2(n-3)/(n-1)}\biggl),\label{p-potential}\\
m\equiv &\frac{4(n-2)V_{n-2}^{(k)}}{(n-1)^2\kappa_n^2}. \label{p-mass}
\end{align} 
Note that since the exponent in the second term is less than 2 for all values of $n$, the potential $V(x)$ is bounded below.
In terms of $x$, the inner product is simply 
\begin{align}
\langle \Psi|\Phi\rangle:=\int^\infty_0\Psi^*\Phi\D{x}.
\end{align}  
We now show that this Hamiltonian operator ${\hat H}$ has the self-adjoint extensions on the half-line $x\geq 0$.

\subsection{Self-adjointness of the Hamiltonian operator}
In order for the quantum system to be well-defined, the evolution must be unitary:
\begin{align}
\frac{\partial}{\partial t}\langle \Psi|\Phi\rangle=0\,.
\end{align}  
This requires the Hamiltonian operator ${\hat H}$ to be self-adjoint in the domain of $0<x<\infty$ or have a self-adjoint extension, which in turn guarantees that the eigenvalues are real. See \cite{robin bcs} for general discussions of self-adjoint extensions of operators.

Using the fall-off condition of $\Psi$ at $x\to \infty$\footnote{This condition is satisfied in the cases we treat in the present paper}, a straightforward calculation reveals that: 
\begin{align}
\frac{\partial}{\partial t}\langle \Psi|\Phi\rangle=\frac{i\hbar}{2m} \biggl(\frac{\partial\Psi^*}{\partial x}\Phi-\Psi^*\frac{\partial\Phi}{\partial x}\biggl)\biggl|_{x=0},
\end{align}  
where we used Eq.~(\ref{Schro}).
Unitarity therefore requires the right-hand side of the above to vanish for all states in the Hilbert space. The wave functions must therefore obey a boundary condition of the form:
\begin{align}
\Psi(0)+L\frac{\partial\Psi}{\partial x}(0)=0,
\label{eq:robin bc}
\end{align}  
where $L$ is a real constant. This corresponds to Dirichlet boundary conditions when $L=0$ and Neumann boundary conditions when $L=\infty$. The remaining values of $L$ give the general Robin boundary conditions \cite{robin bcs}. 

The result (\ref{eq:robin bc}) of the above simplified derivation is valid for the systems considered here. A more general analysis requires the study of the deficiency indices $n_\pm$ which are the dimensions of the space of solutions to the eigenvalue equations:
\begin{align}
{\hat H}\Psi(x) =\pm i\Psi(x)
\end{align}  
without imposing boundary conditions at $x=0$.
The general solutions to the above differential equations contain several parameters for both the upper and lower signs.
The number of parameters, which gives the dimension of the solution space for $\pm i$, is denoted by $n_\pm$.
If $n_\pm=0$, then ${\hat H}$ is essentially self-adjoint and no further boundary conditions are required.
If $n_+\ne n_-$, there are no self-adjoint extensions for ${\hat H}$ and the quantum system is not well-defined.
If $n_+=n_-\ne 0$, then ${\hat H}$ has self-adjoint extensions which require the imposition of further boundary conditions with $n_+=n_-$ parameters. 

It can be verified that for the class of Hamiltonians we consider $n_+ = n_- =1$, so that there is exactly one self-adjoint extension parameter, namely $L$ given above. For all values of this parameter the energy is bounded below, as shown in Appendix~\ref{section:energy bound}. The spectra for different values of $L$ differ only for the low-lying states and must agree in the semi-classical limit. The self-adjoint extension parameter will therefore only be relevant to the three cases below for which we are able to solve the eigenvalue problem exactly. One of these turns out to reduce to the simple harmonic oscillator quantized on the half line. In Appendix~\ref{appendix:half-line}, we review quantization of this system as a specific example of how the energy spectrum depends on the boundary condition at the origin.

%======================================%
%<<<<<<<<<<<< SECTION I  >>>>>>>>>>>>>>%
%======================================%
\section{Exact mass spectrum}
\label{sec:exact}
We must now solve the Schr\"odinger equation (\ref{eq:se 1}) to obtain the mass-energy spectrum.
It is remarkable that one can obtain exact analytic solutions in three specific cases, which we now discuss in turn.

\subsection{BTZ black holes in three dimensions}
In three dimensions ($n=3$), $k=0$ holds and the black hole configuration is possible only for $\Lambda<0$.
In this case, the Hamiltonian operator becomes very simple:
\begin{align}
{\hat H} =\frac{V_{1}^{(0)}}{2\kappa_3^2}\biggl[\frac{x^2}{l^2}-\hbar^2\biggl(\frac{\kappa_3^2}{V_{1}^{(0)}}\biggl)^2\frac{d^2}{dx^2}\biggl],
\end{align}  
which is equivalent to the Hamiltonian of the harmonic oscillator on the half line. A key difference from the usual harmonic oscillator is that in the present case the Hamiltonian is not essentially self-adjoint. One needs to define a self-adjoint extension by imposing suitable boundary conditions at $x=0$. A detailed calculation is provided in
Appendix~\ref{appendix:half-line}.

Defining
\begin{align}
\xi:=\sqrt{\frac{V_{1}^{(0)}}{\hbar l\kappa_3^2}}x,
\end{align}  
we write the Schr\"odinger equation ${\hat H} \psi=E\psi$ as
\begin{align}
\biggl(-\frac{d^2}{d\xi^2}+\xi^{2}\biggl)\psi=&\lambda \psi, \label{harmonic}\\
\lambda:=&\frac{2l}{\hbar}E.
\label{eq:lambda}
\end{align}

If $0\leq \xi <\infty$, then we need to specify a boundary condition at the origin that preserve probability,  rule out imaginary eigenvalues and still leave a non-trivial solution space. 
The family of suitable boundary conditions is:
\be
\psi(0) - \tan\chi \psi'(0) = 0
\ee
with $\chi\in [0,\pi)$, where a prime denote the derivative with respect to $\xi$. 

As shown in
Appendix~\ref{appendix:half-line}, the energy eigenvalues with the symmetric (Neumann) boundary condition $\chi=\pi/2$ is 
\be
E= \frac{\hbar}{l}\biggl(2N + \frac12\biggl)=:E_{\rm s}\quad (N=0,1,2,\cdots).
\ee
Similarly $\chi =0$ gives Dirichlet boundary conditions and the anti-symmetric eigenvalues, namely:
\be
E= \frac{\hbar}{l}\biggl(2N+\frac32\biggl)=:E_{\rm a}.
\ee
In these case, the energy is equally spaced with the unit $2\hbar/l$.
However, this is not the case for general $\chi$, in which one finds a spectrum that is heuristically of the form:
\be
E= \frac{\hbar}{l}\biggl(2N + \frac{1}{2} + \alpha_N(\chi)\biggl)=:E_\chi
\ee
where $\alpha_N(\chi)$ depends slightly on $N$ and hence $E$ is not equally spaced. In the large $N$ limit the first term dominates and the spectrum goes to
\be
E\to_{N\to\infty}\frac{2\hbar N}{l}
\ee
which is independent of the extension parameter as expected.

It is interesting to note that for a finite range of $\tan{\chi}$ the ground state energy $E$ is negative. In order for the system to describe  a black hole, the mass, and hence energy, must be positive, so that these negative energy ground states do not have a classical analogue. This is similar to the situation for the free particle on the half line~\cite{robin bcs}. For a range of values of the extension parameter there exists a unique negative energy bound state in addition to the expected positive energy scattering states. This bound state again has no classical analogue.
It is important to keep in mind, each value of $\chi$ corresponds to a different, inequivalent quantization and there are no transitions between different boundary conditions are allowed.

%%%%%%%%%%%%%%%%%OTHER BTZ Calculations%%%%%%%%%%%%%%%%%%%%%%%%%%%%%%%%
We close this subsection by recalling that the mass spectrum of the BTZ black hole has been obtained by many other methods. As discussed in greater detail in the conclusions, Ref.~\cite{carlip} used semi-classical arguments based on the exactly known quasinormal mode spectrum of BTZ black holes, with results that coincide with ours for Dirichlet and Neumann boundary conditions. As pointed out in \cite{carlip} this spectrum also agrees with what one would expect from the AdS/CFT correspondence by examining the dual operators in the CFT. 

Other papers obtain differing results. For example, in \cite{Setare2004} an equally spaced mass spectrum is obtained, with spacing half of our value. The analysis is similar to that of Carlip, and the extraneous factor of 2 is a consequence of neglecting the fact that the quasinormal modes separate into left and right movers, each of which carry non-zero angular momentum quantum numbers. Zero angular momentum excitations require one of each type, so that the spacing for zero angular momentum black holes is double that derived in \cite{Setare2004}, in agreement with our results.
In \cite{vgksw2008}, the spectrum is derived using the Wheeler-Dewitt equation. The result is again that of a simple harmonic oscillator $M=\hbar/l(N+1/2)$ without the factor of 2. The discrepancy here is likely due the need for a careful treatment of the boundary conditions which leads to the need for self-adjoint extensions which give rise to distinct sectors.  Dirichlet boundary conditions yield the anti-symmetric states of the oscillator, while the Neumann boundary conditions yield the symmetric states. These states live in different Hilbert spaces.  The correct spacing is therefore double that obtained in [3]. Finally, we mention Ref.~\cite{kn2010} where an equally spaced area spectrum is obtained based on the Bohr correspondence principle applied to the imaginary part of the highly damped quasinormal modes, as advocated by Maggiore~\cite{Maggiore2008}. The mass spectrum therefore goes like the $\sqrt{N}$ in disagreement with ours. We must conclude that at least in the context of the BTZ black hole, for which the analytic form of the quasinormal modes are known exactly so that it is not necessary to go to the highly damped limit, Maggiore's arguments do not apply.

\subsection{Toroidal black holes in arbitrary dimensions}
The Schr\"odinger equation with $k=0$ in arbitrary $n$ also reduces to Eq.~(\ref{harmonic}), where $\xi$ and $\lambda$ are now
\begin{align}
\xi:=&\sqrt{\frac{2(n-2)V_{n-2}^{(0)}}{(n-1)\hbar l\kappa_n^2}}x,\\
\lambda:=&\frac{4l}{(n-1)\hbar}E.
\end{align}  
In this case, $\Lambda<0$ is required for black-hole configurations to exist. Moreover, the domain of $\xi$ is $0\le \xi<\infty$.
The spectrum of the toroidal black hole is therefore similar to the one given above for the BTZ black hole. Again, the energy spectrum of this toroidal black hole with the Dirichlet boundary condition is 
\begin{align}
E=\frac{(n-1)\hbar}{2l}\biggl(2N+\frac12\biggl) \quad (N=0,1,2,\cdots). \label{EN:toroidal}
\end{align}

\subsection{Five-dimensional black holes}
In five dimensions ($n=5$), the Hamiltonian operator simplifies a great deal: 
\begin{align}
{\hat H} =\frac{3V_{3}^{(k)}}{2\kappa_5^2}\biggl[\frac{x^2}{l^2}+kx-\hbar^2\biggl(\frac{2\kappa_5^2}{3V_{3}^{(k)}}\biggl)^2\frac{d^2}{dx^2}\biggl].
\end{align}  
The general solution to Schr\"odinger equation ${\hat H}\psi=E\psi$ can again be found analytically.

\subsubsection{$\Lambda<0$}
We consider the case of $\Lambda<0$ with $k=\pm 1$ because the case with $k=0$ was studied in the previous subsection.
In this case, the Schr\"odinger equation reduces to Eq.~(\ref{harmonic}), where $\lambda$ and $\xi$ are now defined by 
\begin{align}
\lambda:=&\frac{l}{\hbar}\biggl(E+\frac{3k^2l^2V_{3}^{(k)}}{8\kappa_5^2}\biggl),\\
\xi:=&\sqrt{\frac{3V_{3}^{(k)}}{2\hbar l\kappa_5^2}}\biggl(x+\frac{kl^2}{2}\biggl).
\end{align}  
{ Although it is possible to solve the Schr\"odinger equation, the change of variables required to put the Hamiltonian in simple harmonic form makes it more difficult to obtain the exact mass spectrum in this case.}
The difference from the toroidal case is the value of $\xi$ at the boundary $x=0$.
We have to impose the boundary condition at 
\begin{align}
\xi=\frac{kl^2}{2}\sqrt{\frac{3V_{3}^{(k)}}{2\hbar l\kappa_5^2}}=:\xi_0.
\end{align}  
The boundary condition is 
\begin{align}
\psi(\xi_0)+L\psi'(\xi_0)=0,
\end{align}  
where $L$ is a constant.
%The Dirichlet boundary condition with $L=0$ is 
%\begin{align}
%\psi(\xi_0)=0.
%\end{align}  
%For simplicity we therefore relegate this case to the next section where we obtain the mass spectrum for large $E$ in the WKB approximation.

\subsubsection{$\Lambda=0$}

In the absence of $\Lambda$, $k=1$ is required for the black-hole configurations to exist.
Defining 
\begin{align}
\xi:=\biggl(\frac{3V_{3}^{(1)}}{2\hbar\kappa_5^2}\biggl)^{2/3}x,
\end{align}  
we write the Schr\"odinger equation it in the following form:
\begin{align}
&\biggl(-\frac{d^2}{d\xi^2}+\xi\biggl)\psi=\lambda\psi,\\
&\lambda:=\biggl(\frac{2\kappa_5^2}{3\hbar^2V_{3}^{(1)}}\biggl)^{1/3}E.
\end{align}  
The general solution of this equation is given in terms of the Airy functions:
\begin{align}
\psi(\xi)=F_1\mbox{Ai}(\xi-\lambda)+F_2\mbox{Bi}(\xi-\lambda),
\end{align}  
where $F_1$ and $F_2$ are constants and 
\begin{align}
\mbox{Ai}(z):=&\frac{1}{\pi}\int^\infty_0\cos\biggl(\frac{t^3}{3}+zt\biggl)dt,\\ 
\mbox{Bi}(z):=&\frac{1}{\pi}\int^\infty_0\biggl[\sin\biggl(\frac{t^3}{3}+zt\biggl)+\exp\biggl(-\frac{t^3}{3}+zt\biggl)\biggl]dt.
\end{align}  

Since $\mbox{Bi}(z)$ is diverging for $z\to \infty$, the normalization for $\psi$ requires $F_2=0$.
The asymptotic expansion of $\psi$ around $\xi=0$ is given by 
\begin{align}
\psi(\xi)\simeq F_1\mbox{Ai}(-\lambda)+F_1\frac{d\mbox{Ai}}{d\xi}\biggl|_{\xi=0}\xi,
\end{align}  
which shows 
\begin{align}
\psi(0)=&F_1\mbox{Ai}(-\lambda),\\
\psi'(0)=&F_1\frac{d\mbox{Ai}}{d\xi}\biggl|_{\xi=0}.
\end{align}  

The boundary conditions implied by (\ref{eq:robin bc}) for arbitrary extension parameter $L$ are therefore:
\be
\mbox{Ai}(-\lambda)+L \frac{d\mbox{Ai}}{d\xi}\biggl|_{\xi=0}=0
\ee
It is known that the zeros of $\mbox{Ai}(z)=0$ for large negative $z$ are given by 
\begin{align}
z\simeq -\biggl(\frac{3\pi(4K-1)}{8}\biggl)^{2/3},
\end{align}  
while the zeros of $d\mbox{Ai}/dy(z)=0$ for large negative $z$ are:
\begin{align}
z\simeq -\biggl(\frac{3\pi(4K-3)}{8}\biggl)^{2/3},
\end{align}
where $K$ is a large positive integer.  
Therefore in the large $K$ limit, the spectrum for arbitrary extension parameter goes to 
\begin{align}
E\simeq  \frac32\biggl(\frac{\pi^2\hbar^2V_{3}^{(1)}}{\kappa_5^2}\biggl)^{1/3}K^{2/3}. \label{E-five1}
\end{align}

The result (\ref{E-five1}) shows that the mass spectrum of the black hole is not equally spaced.
In this case, Eq.~(\ref{entropy}) shows that the entropy $S$ of the black hole is equally spaced for large $N$:
\begin{align}
S \simeq 2\pi^2\hbar N=\pi h N. \label{S5}
\end{align}  
This is a sharp difference from the asymptotically AdS case, in which not the entropy but the mass of the black hole is equally spaced.

%======================================%
%<<<<<<<<<<<< SECTION I  >>>>>>>>>>>>>>%
%======================================%
\section{Mass spectrum in the WKB approximation}
\label{sec:WKB}

Although we considered the cases in which the Schr\"odinger equation is solved analytically in the last section, it is not possible in general.
In this section, we obtain the mass spectrum of the black hole for large $E$ using the WKB approximation.

The Schr\"odinger equation is written as
\begin{align}
\biggl(-\frac{\hbar^2}{2m}\frac{d^2}{dx^2}+V(x)\biggl)\psi=E\psi, \label{sch1}
\end{align}  
where the potential $V(x)$ and the mass $m$ are defined by Eqs.~(\ref{p-potential}) and (\ref{p-mass}), respectively.
The general solution of this differential equation around $x=0$ is 
\begin{align}
\psi(x)\simeq &A\sin\sqrt{\bar E}x+B\cos\sqrt{\bar
E}x=\sqrt{A^2+B^2}\sin\left(\sqrt{\bar E}x+\phi_0\right),\label{sol:x=0}\\
{\bar
E}:=&\frac{2mE}{\hbar^2}=\frac{8(n-2)V_{n-2}^{(k)}E}{(n-1)^2\hbar^2\kappa_n^2},
\label{eq:}
\end{align}
where $A$ and $B$ are constants and $\phi_0:=\arctan(B/A)$.

We will consider the WKB solution which is valid away from the turning point $x=a$.
Connecting the WKB solution to the asymptotic solution (\ref{sol:x=0}), we will obtain the mass/energy spectrum.
The turning point $x=a$ is defined by 
\begin{align}
E-V(a)=0.
\end{align}  
The WKB solution for $x<a$ is  
\begin{align}
\psi(x) \simeq&\frac{2D_-}{\sqrt{k(x)}}\sin\biggl(\eta(x,a)+\frac{\pi}{4}\biggl), \label{WKBsolution1} \\
k(x):=&\frac{\sqrt{2m(E-V(x))}}{\hbar},\\
\eta(x,a):=&\int^a_x \frac{\sqrt{2m|E-V({\bar x})|}}{\hbar}d{\bar x}=\int^a_x \frac{\sqrt{2m(E-V({\bar x}))}}{\hbar}d{\bar x}, \label{eta}
\end{align}  
which connects to the following normalizable WKB solution in the region of $x>a$: 
\begin{align}
\psi(x) \simeq \frac{D_-}{\sqrt{\kappa(x)}}e^{-\eta(a,x)}. \label{WKBsolution2}
\end{align}  

In some cases, we are able to obtain $\eta(x,a)$ in a closed form but not in general.
In such a case, using the following the asymptotic behaviors of $k(x)$ and $\eta(x,a)$ around $x=0$;
\begin{align}
k(0)=&\frac{\sqrt{2mE}}{\hbar},\\
\eta(x,a)\simeq &\eta(0,a)-\frac{\sqrt{2m|E-V(0)|}}{\hbar}x \nonumber \\
=&\eta(0,a)-\frac{\sqrt{2m E}}{\hbar}x,
\end{align}  
we obtain the WKB solution (\ref{WKBsolution1}) around $x=0$ as
\begin{align}
\psi(x) \simeq&\frac{2D_-}{\sqrt{k(0)}}\sin\biggl(-\sqrt{{\bar E}}x+\eta(0,a)+\frac{\pi}{4}\biggl),
\end{align}  
which is compared with Eq.~(\ref{sol:x=0}) in order to obtain the energy spectrum. The task then is to compute $\eta(0,a)$ and then apply the boundary conditions (\ref{eq:robin bc}).

\subsection{Asymptotically flat black holes}
Here let us consider the asymptotically flat spherically symmetric black hole, namely $k=1$ and $\Lambda=0$.
We have
\begin{align}
E-V(x)=E-\frac{(n-2)V_{n-2}^{(1)}}{2\kappa_n^2}x^{2(n-3)/(n-1)}
\end{align}  
and so the turning point $x=a$ is given by 
\begin{align}
a^{2(n-3)/(n-1)}=\frac{2\kappa_n^2E}{(n-2)V_{n-2}^{(1)}}.
\end{align}  

Let us now consider the approximate expression of the WKB solution (\ref{WKBsolution1}) for large $E$ using
\begin{align}
k(x)=\frac{\sqrt{2m}}{\hbar}\biggl(E-\frac{(n-2)V_{n-2}^{(1)}}{2\kappa_n^2}x^{2(n-3)/(n-1)}\biggl)^{1/2}
\end{align}  
and 
\begin{align}
\eta(x,a)=\frac{2}{(n-3)\hbar}\biggl(\frac{2\kappa_n^2}{(n-2)V_{n-2}^{(1)}}\biggl)^{1/(n-3)}\int^E_{\bar b} b^{-(n-5)/2(n-3)}\sqrt{E-b}db,\label{int-key-flat}
\end{align}  
where we defined
\begin{align}
b:=&\frac{(n-2)V_{n-2}^{(1)}}{2\kappa_n^2}{\bar x}^{2(n-3)/(n-1)}.\\
{\bar b}:=&\frac{(n-2)V_{n-2}^{(1)}}{2\kappa_n^2}x^{2(n-3)/(n-1)}. \label{bbar-flat}
\end{align}  
The integral (\ref{int-key-flat}) is computed analytically for $n=4,5$, but not for $n\ge 6$.
We discuss these cases separately.

\subsubsection{$n=4$}
For $n=4$, we obtain $\eta(x,a)$ in a closed form:
\begin{align}
\eta(x,a)=&\frac{2}{\hbar}\frac{\kappa_4^2}{V_{2}^{(1)}}\int^E_{\bar b} b^{1/2}\sqrt{E-b}db \nonumber \\
=&\frac{2}{\hbar}\frac{\kappa_4^2}{V_{2}^{(1)}}\biggl(\frac{\pi E^2}{16}+\frac{{\bar b}^{1/2}}{2}(E-{\bar b} )^{3/2}-\frac{E}{4}\sqrt{{\bar b} (E-{\bar b} )}-\frac{E^2}{8}\arctan\frac{{\bar b} -E/2}{\sqrt{{\bar b} (E-{\bar b} )}}\biggl).
\end{align}  
For large $E$, this reduces to
\begin{align}
\eta(x,a)\simeq \frac{\kappa_4^2}{V_{2}^{(1)}}\frac{\pi E^2}{4\hbar}-\biggl(\frac{16V_{2}^{(1)}E}{9\hbar^2\kappa_4^2}\biggl)^{1/2}x,
\end{align}  
where we used Eq.~(\ref{bbar-flat}) and 
\begin{align}
&\arctan\frac{{\bar b} -E/2}{\sqrt{{\bar b} (E-{\bar b} )}}\simeq -\frac{\pi}{2}+2\sqrt{\frac{{\bar b}}{E}}+\frac13\biggl(\frac{{\bar b}}{E}\biggl)^{3/2}.
\end{align}  
From Eq.~(\ref{WKBsolution1}), we obtain the WKB solution for large $E$:
\begin{align}
\psi \simeq &2D_-\biggl(\frac{\hbar}{\sqrt{2m E}}\biggl)^{1/2}\sin\biggl(\frac{\kappa_4^2}{V_{2}^{(1)}}\frac{\pi E^2}{4\hbar}-\biggl(\frac{16V_{2}^{(1)}E}{9\hbar^2\kappa_4^2}\biggl)^{1/2}x+\frac{\pi}{4}\biggl).  \label{WKBlast}
\end{align}  

We compare the above the WKB solution (\ref{WKBlast}) with the solution (\ref{sol:x=0}).
Imposing the boundary conditions (\ref{eq:robin bc}) one finds the following condition for large $E$, again independent of the extension parameter:
\begin{align}
&-\frac{\kappa_4^2}{V_{2}^{(1)}}\frac{\pi E^2}{4\hbar}-\frac{\pi}{4}=-\pi N,   \\
\to &E^2 \simeq \frac{4\hbar V_{2}^{(1)}}{\kappa_4^2}N, \label{E-flat}
\end{align}  
where $N$ is a positive integer.

Our result (\ref{E-flat}) coincides with the result by Louko and M\"akel\"a and shows that the mass of the black hole is not equally spaced for large $E$.
In contrast, using Eq.~(\ref{entropy}), we show that the entropy of the black hole is equally spaced:
\begin{align}
S\simeq 8\pi \hbar N.
\end{align}

\subsubsection{$n=5$}
In the case of $n=5$, the analysis is much simpler.
We obtain
\begin{align}
\eta(x,a)=&\frac{1}{\hbar}\biggl(\frac{2\kappa_5^2}{3V_{3}^{(1)}}\biggl)^{1/2}\int^E_{\bar b}\sqrt{E-b}db \nonumber \\
=&\frac{1}{\hbar}\biggl(\frac{2\kappa_5^2}{3V_{3}^{(1)}}\biggl)^{1/2}\frac23\biggl(E-\frac{3V_{3}^{(1)}}{2\kappa_5^2}x\biggl)^{3/2}
\end{align}  
and Eq.~(\ref{WKBsolution1}) becomes
\begin{align}
\psi(x) \simeq 2D_-\biggl(\frac{\hbar}{\sqrt{2m E}}\biggl)^{1/2}\sin\biggl(\biggl(\frac{8\kappa_5^2}{27\hbar^2V_{3}^{(1)}}\biggl)^{1/2}E^{3/2}-\biggl(\frac{3V_{3}^{(1)}E}{2\hbar^2\kappa_5^2}\biggl)^{1/2}x+\frac{\pi}{4}\biggl).  \label{WKBlast:n=5}
\end{align}  

Comparing the above the WKB solution (\ref{WKBlast:n=5}) with the solution (\ref{sol:x=0}) and imposing the boundary condition (\ref{eq:robin bc}) we obtain the spectrum for large $E$ as
\begin{align}
&-\biggl(\frac{8\kappa_5^2}{27\hbar^2V_{3}^{(1)}}\biggl)^{1/2}E^{3/2}-\frac{\pi}{4}=-\pi N, \nonumber \\
\to &E^{3/2} \simeq \biggl(\frac{27\hbar^2V_{3}^{(1)}}{8\kappa_5^2}\biggl)^{1/2}\pi N,
\end{align}  
where $N$ is an integer.
Using Eq.~(\ref{entropy}), we show that the entropy of the black hole is again equally spaced for large $E$:
\begin{align}
S\simeq 2\pi^2 \hbar  N.
\end{align}
This coincides with Eq.~(\ref{S5}).

\subsubsection{$n \ge 6$}
For $n\ge 6$, the nontrivial task is how we evaluate the integral (\ref{int-key-flat}) for large $E$.
The expression of $\eta(x,a)$ around $x=0$ is given by  
\begin{align}
\eta(x,a)\simeq&\eta(0,a)+\frac{d{\bar b}}{dx}\frac{d\eta(x,a)}{d{\bar b}}\biggl|_{x=0}x \nonumber \\
=&\eta(0,a)-\sqrt{\frac{8(n-2)V_{n-2}^{(1)}E}{(n-1)^2\hbar^2\kappa_n^2}}x.
\end{align}  
Let us obtain $\eta(0,a)$.
We first compute
\begin{align}
\eta(0,a)=&\frac{2}{(n-3)\hbar}\biggl(\frac{2\kappa_n^2}{(n-2)V_{n-2}^{(1)}}\biggl)^{1/(n-3)}\int^E_0 E^{1/2}b^{-(n-5)/2(n-3)}\sqrt{1-b/E}db \nonumber \\
=&\frac{4E^{(n-2)/(n-3)}}{(n-1)\hbar}\biggl(\frac{2\kappa_n^2}{(n-2)V_{n-2}^{(1)}}\biggl)^{1/(n-3)}\int^1_0\sqrt{1-\zeta^{2(n-3)/(n-1)}}d\zeta \nonumber \\
=&\frac{2E^{(n-2)/(n-3)}}{(n-3)\hbar}\biggl(\frac{2\kappa_n^2}{(n-2)V_{n-2}^{(1)}}\biggl)^{1/(n-3)}B\biggl(\frac{n-1}{2(n-3)},\frac32\biggl),
\end{align}  
where we defined
\begin{align}
\zeta:=&\biggl(\frac{b}{E}\biggl)^{(n-1)/2(n-3)}
\end{align}  
and $B(x,y)$ is the Beta function.
The last equality is shown as
\begin{align}
B(x,y):=&\int_0^1t^{x-1}(1-t)^{y-1}dt \nonumber \\
=&\frac{2(n-3)}{n-1}\int_0^1\eta^{[2(n-3)(x-1)+(n-5)]/(n-1)}(1-\eta^{2(n-3)/(n-1)})^{y-1}d\eta,\\
\to B\biggl(\frac{n-1}{2(n-3)},\frac32\biggl)=&\frac{2(n-3)}{n-1}\int_0^1(1-\eta^{2(n-3)/(n-1)})^{1/2}d\eta,
\end{align}  
where we used
\begin{align}
\eta:=t^{(n-1)/2(n-3)}.
\end{align}  
The expression of $\eta(0,a)$ in terms of the Gamma function $\Gamma(x)$ is 
\begin{align}
\eta(0,a)=&\frac{2E^{(n-2)/(n-3)}}{(n-3)\hbar}\biggl(\frac{2\kappa_n^2}{(n-2)V_{n-2}^{(1)}}\biggl)^{1/(n-3)}\frac{\Gamma(\frac{n-1}{2(n-3)})\Gamma(\frac32)}{\Gamma(\frac{n-1}{2(n-3)}+\frac32)} \nonumber \\
=&\frac{\sqrt{\pi}E^{(n-2)/(n-3)}}{(n-2)\hbar}\biggl(\frac{2\kappa_n^2}{(n-2)V_{n-2}^{(1)}}\biggl)^{1/(n-3)}\frac{\Gamma(\frac{n-1}{2(n-3)})}{\Gamma(\frac{n-1}{2(n-3)}+\frac12)}.
\end{align}

The WKB solution for $x<a$ behaves around $x=0$ as  
\begin{align}
\psi(x) \simeq&\frac{2D_-}{\sqrt{k(x)}}\sin\biggl(\eta(x,a)+\frac{\pi}{4}\biggl)  \nonumber \\
\propto &\sin\biggl(\eta(0,a)-\sqrt{\frac{8(n-2)V_{n-2}^{(1)}E}{(n-1)^2\hbar^2\kappa_n^2}}x+\frac{\pi}{4}\biggl)  \label{WKBsol:n6}
\end{align}  
Comparing the above the WKB solution (\ref{WKBsol:n6}) with the solution (\ref{sol:x=0}) and imposing the Robin boundary conditions at $x=0$, we obtain the spectrum for large $E$ to be;
\begin{align}
&-\eta(0,a)-\frac{\pi}{4}=-\pi N \nonumber \\
\to &E^{(n-2)/(n-3)} \simeq (n-2)\hbar\sqrt{\pi} N\frac{\Gamma(\frac{n-1}{2(n-3)}+\frac12)}{\Gamma(\frac{n-1}{2(n-3)})}\biggl(\frac{(n-2)V_{n-2}^{(1)}}{2\kappa_n^2}\biggl)^{1/(n-3)},  \label{E-n}
\end{align}  
where $N$ is a positive integer.

Using Eq.~(\ref{entropy}), we show that the entropy of the black hole is equally spaced for large $E$:
\begin{align}
 S\simeq  &4\pi^{3/2}\hbar \frac{\Gamma(\frac{n-1}{2(n-3)}+\frac12)}{\Gamma(\frac{n-1}{2(n-3)})}N=:S_N.
\end{align}
Some particular cases are given explicitly as
\begin{eqnarray}
S_N=
\left\{ \begin{array}{ll}
4\pi^{3/2}\hbar N\frac{\Gamma(2)}{\Gamma(3/2)}=8\pi \hbar N~~(n=4)\\
4\pi^{3/2}\hbar N\frac{\Gamma(3/2)}{\Gamma(1)} =2\pi^2 \hbar  N~(n=5)\\
4\pi^{3/2}\hbar \frac{\Gamma(1)}{\Gamma(1/2)}N=4\pi\hbar N~~(n\to \infty),
\end{array} \right.
\end{eqnarray}
which is consistent with the results in the previous subsections.

\subsection{Asymptotically anti-de~Sitter black holes}
Next let us consider asymptotically AdS black holes.
Since the mass spectrum of the toroidal black hole ($k=0$) was already derived, we consider only the case with $k=\pm 1$ and $n\ge 4$ in this subsection.
Since the function $\eta(x,a)$ defined by Eq.~(\ref{eta}) is completed analytically for $n=5$, we treat $n=5$ and $n\ne  5$ separately.

\subsubsection{$k=\pm 1$ with $n=5$}
Here we consider the case of $n=5$ with $k=\pm 1$.
We have
\begin{align}
E-V(x)=E-\frac{3V_{n-2}^{(k)}}{2\kappa_n^2}\biggl(\frac{x^2}{l^2}+kx\biggl)
\end{align}  
and so the turning point $x=a$ is determined by 
\begin{align}
E=\frac{3V_{n-2}^{(k)}}{2\kappa_n^2}\biggl(\frac{a^2}{l^2}+ka\biggl).
\end{align}  
The WKB analysis is valid for large $E$, where there is only one turning point.

We consider the approximate expression of the WKB solution (\ref{WKBsolution1}) for large $E$ and compare with the solution (\ref{sol:x=0}) in order to obtain the energy spectrum.
In the present case, $k(x)$ and $\eta(x,a)$ are given as
\begin{align}
k(x)=\frac{\sqrt{2m}}{\hbar}\biggl(E-\frac{3V_{n-2}^{(k)}}{2\kappa_n^2}(l^{-2}a^2+ka)\biggl)^{1/2}
\end{align}  
and 
\begin{align}
\eta(x,a)= & \frac{\sqrt{2m E}}{\hbar} \biggl[\frac12\sqrt{\frac{l^2}{p}}\biggl(1+\frac{pl^2k^2}{4}\biggl)\frac{\pi}{2} -\frac12\biggl(x+\frac{kl^2}{2}\biggl)\sqrt{1-\frac{px^2}{l^2}-pkx} \nonumber \\
&-\frac12\sqrt{\frac{l^2}{p}}\biggl(1+\frac{pl^2k^2}{4}\biggl)\arctan\biggl\{\sqrt{\frac{p/l^2}{1-px^2/l^2-pkx}}\biggl(x+\frac{kl^2}{2}\biggl)\biggl\}\biggl],  \\
p:= &\frac{3V_{n-2}^{(k)}}{2\kappa_n^2E},
\end{align} 
respectively, where we used $1=p(l^{-2}a^2+ka)$ and the fact $a+kl^2/4>0$.
For large $E$, $\eta(x,a)$ becomes
\begin{align}
\eta(x,a)\simeq -\sqrt{\frac{3V_{n-2}^{(k)}E}{2\hbar^2\kappa_n^2}}x+\frac{\pi l}{4\hbar}E. \label{WKBsolution1:k=0}
\end{align} 

Comparing the above the WKB solution (\ref{WKBsolution1:k=0}) with the solution (\ref{sol:x=0}) with the Dirichlet boundary condition $B=0$ at $x=0$, we obtain the spectrum for large $E$ as
\begin{align}
&\frac{\pi l}{4\hbar}E+\frac{\pi}{4}=\pi N \nonumber \\
\to &  E \simeq \frac{4\hbar}{l}N, \label{E-AdS5}
\end{align}  
where $N$ is a positive integer.
As in the previous cases, the large $N$ spectrum is independent of the boundary conditions at the origin.

\subsubsection{$k=\pm 1$ with $n=4$ or $n\ge 6$}

For  $n=4$ or $n\ge 6$, we have
\begin{align}
k(x)=\frac{\sqrt{2m}}{\hbar}\biggl(E-\frac{(n-2)V_{n-2}^{(k)}}{2\kappa_n^2}\biggl(\frac{x^2}{l^2}+kx^{2(n-3)/(n-1)}\biggl)\biggl)^{1/2}
\end{align}  
and 
\begin{align}
\eta(x,a)=&\int^a_x \frac{\sqrt{2m}}{\hbar}\biggl(E-\frac{(n-2)V_{n-2}^{(k)}}{2\kappa_n^2}\biggl(\frac{x^2}{l^2}+kx^{2(n-3)/(n-1)}\biggl)\biggl)^{1/2}d{\bar x} \label{eta-x}  \\
=&\frac{2}{(n-3)\hbar}\biggl(\frac{2\kappa_n^2}{(n-2)V_{n-2}^{(k)}}\biggl)^{1/(n-3)} \nonumber \\
&\times \int^{{\bar a}(a)}_{{\bar b}(x)} b^{-(n-5)/2(n-3)}\sqrt{E-kb-\frac{1}{l^2}\biggl(\frac{2\kappa_n^2}{(n-2)V_{n-2}^{(k)}}\biggl)^{2/(n-3)}b^{(n-1)/(n-3)}}db,\label{int-key}
\end{align}  
where we defined
\begin{align}
b:=&\frac{(n-2)V_{n-2}^{(k)}}{2\kappa_n^2}{\bar x}^{2(n-3)/(n-1)}.\\
{\bar b}:=&\frac{(n-2)V_{n-2}^{(k)}}{2\kappa_n^2}x^{2(n-3)/(n-1)} \label{bbar}
\end{align}  
and ${\bar a}$ is determined by
\begin{align}
0=E-k{\bar a}+{\tilde \Lambda}\biggl(\frac{2\kappa_n^2}{(n-2)V_{n-2}^{(k)}}\biggl)^{2/(n-3)}{\bar a}^{(n-1)/(n-3)}.
\end{align}

Since the integral (\ref{int-key}) is not obtained in a closed form for $n\ge 6$ with $k=\pm 1$, we will evaluate this integral around $x=0$.
Near $x=0$, $\eta(x,a)$ is approximated by
\begin{align}
\eta(x,a)\simeq&\eta(0,a)+\frac{d{\bar b}}{dx}\frac{d\eta(x,a)}{d{\bar b}}\biggl|_{x=0}x \nonumber \\
=&\eta(0,a)-\sqrt{\frac{8(n-2)V_{n-2}^{(k)}E}{(n-1)^2\hbar^2\kappa_n^2}}x, \label{eta-x0}
\end{align}  
where we used
\begin{align}
\frac{d\eta(x,a)}{d{\bar b}}=&-\frac{2}{(n-3)\hbar}\biggl(\frac{2\kappa_n^2}{(n-2)V_{n-2}^{(k)}}\biggl)^{1/(n-3)} \nonumber \\
&\times {\bar b}^{-(n-5)/2(n-3)}\sqrt{E-k{\bar b}-\frac{1}{l^2}\biggl(\frac{2\kappa_n^2}{(n-2)V_{n-2}^{(k)}}\biggl)^{2/(n-3)}{\bar b}^{(n-1)/(n-3)}}.
\end{align}

We will evaluate the following integral for large $E$:
\begin{align}
\eta(0,a)=&\frac{2}{(n-3)\hbar}\biggl(\frac{2\kappa_n^2}{(n-2)V_{n-2}^{(k)}}\biggl)^{1/(n-3)} \nonumber \\
&\times \int^{\bar a}_0 b^{-(n-5)/2(n-3)}\sqrt{E-kb+{\tilde \Lambda}\biggl(\frac{2\kappa_n^2}{(n-2)V_{n-2}^{(k)}}\biggl)^{2/(n-3)}b^{(n-1)/(n-3)}}db.
\end{align}  
Defining $y:=b^{(n-1)/2(n-3)}$, $\alpha:=(n-3)/(n-1)$, $l^2=-1/{\tilde \Lambda}$, and $\Pi:=2\kappa_n^2/(n-2)V_{n-2}^{(k)}$, we have
\begin{align}
\eta(0,a)=&\frac{4\Pi^{1/(n-3)} }{(n-1)\hbar} \int^{\bar y}_0 \sqrt{E-ky^{2\alpha}-\frac{\Pi^{2/(n-3)}}{l^2}y^{2}}dy,\\
=&\frac{4\Pi^{1/(n-3)} }{(n-1)\hbar}\int^{\bar y}_0 \sqrt{E-\frac{\Pi^{2/(n-3)}}{l^2}y^{2}}\sqrt{1-\frac{ky^{2\alpha}}{E-(\Pi^{2/(n-3)}/l^2)y^{2}}}dy,
\end{align}  
where ${\bar y}$ is defined by 
\begin{align}
0=E-k{\bar y}^{2\alpha}-\frac{\Pi^{2/(n-3)}}{l^2}{\bar y}^{2}. \label{turn-y}
\end{align}  

For large $E$, the algebraic equation (\ref{turn-y}) is approximated to be 
\begin{align}
0\simeq E-\frac{\Pi^{2/(n-3)}}{l^2}{\bar y}^{2}
\end{align}  
and hence the turning point $y={\bar y}$ is approximated to be 
\begin{align}
{\bar y}^{2}\simeq \frac{El^2}{\Pi^{2/(n-3)}}. \label{ybar-zeroth}
\end{align}  
This is the zeroth-order approximation for ${\bar y}$.

The first-order approximation is given by putting Eq.~(\ref{ybar-zeroth}) into the second term of Eq.~(\ref{turn-y}).
The result is
\begin{align}
{\bar y}^{2}\simeq \frac{El^2}{\Pi^{2/(n-3)}}-\frac{kl^2}{\Pi^{2/(n-3)}}\biggl(\frac{El^2}{\Pi^{2/(n-3)}}\biggl)^{\alpha}. 
\end{align}  
Now we define
\begin{align}
y_0^2:=\frac{El^2}{\Pi^{2/(n-3)}}-d\biggl(\frac{El^2}{\Pi^{2/(n-3)}}\biggl)^{\beta},
\end{align}  
where $d$ is a constant and another constant $\beta$ satisfies $\alpha<\beta<1$.
Then, we consider the following integral:
\begin{align}
{\bar \eta}:=&\frac{4\Pi^{1/(n-3)} }{(n-1)\hbar} \int^{y_0}_0 \sqrt{E-ky^{2\alpha}-\frac{\Pi^{2/(n-3)}}{l^2}y^{2}}dy \\
=&\frac{4\Pi^{1/(n-3)} }{(n-1)\hbar}\int^{y_0}_0 \sqrt{E-\frac{\Pi^{2/(n-3)}}{l^2}y^{2}}\sqrt{1-\frac{ky^{2\alpha}}{E-(\Pi^{2/(n-3)}/l^2)y^{2}}}dy. \label{bareta}
\end{align}  
The difference between $\eta(0,a)$ and ${\bar \eta}$ is the upper bound of the integral.
The second term in the second square root goes to zero as $\mathcal{O}(E^{-1})$ for $E\to \infty$, namely far from $y_0$. 
Close to $y=y_0$, it goes to zero for $E\to \infty$ as
\begin{align}
\frac{ky^{2\alpha}}{E-(\Pi^{2/(n-3)}/l^2)y^{2}}\simeq&ky_0^{2\alpha}\frac{l^2}{d\Pi^{2/(n-3)}}\biggl(\frac{El^2}{\Pi^{2/(n-3)}}\biggl)^{-\beta} \nonumber \\
=&\frac{kl^2}{d\Pi^{2/(n-3)}}\biggl(\frac{El^2}{\Pi^{2/(n-3)}}-d\biggl(\frac{El^2}{\Pi^{2/(n-3)}}\biggl)^{\beta}\biggl)^{\alpha}\biggl(\frac{El^2}{\Pi^{2/(n-3)}}\biggl)^{-\beta} \nonumber \\
\simeq &\frac{kl^2}{d\Pi^{2/(n-3)}}\biggl(\frac{El^2}{\Pi^{2/(n-3)}}\biggl)^{\alpha-\beta}\to 0.
\end{align}  
Hence, for large $E$, ${\bar \eta}$ can be expanded as
\begin{align}
{\bar \eta}\simeq &\frac{4\Pi^{1/(n-3)}}{(n-1)\hbar}\int^{y_0}_0 \sqrt{E-\frac{\Pi^{2/(n-3)}}{l^2}y^{2}}\biggl(1-\frac12 \frac{ky^{2\alpha}}{E-(\Pi^{2/(n-3)}/l^2)y^{2}}\biggl)dy \nonumber \\
=&\frac{4\Pi^{1/(n-3)}}{(n-1)\hbar}\int^{y_0}_0 \sqrt{E-\frac{\Pi^{2/(n-3)}}{l^2}y^{2}}dy-\frac{2\Pi^{1/(n-3)}}{(n-1)\hbar}\int^{y_0}_0 \frac{ky^{2\alpha}}{\sqrt{E-(\Pi^{2/(n-3)}/l^2)y^{2}}}dy.
\end{align}  

Since the above ${\bar \eta}$ approximates $\eta(0,a)$, we evaluate ${\bar \eta}$ for large $E$.
Defining 
\begin{align}
x:=\sqrt{\frac{\Pi^{2/(n-3)}}{El^2}}y,
\end{align}  
we obtain 
\begin{align}
{\bar \eta}\simeq &\frac{4El}{(n-1)\hbar}\int^{x_0}_0 \sqrt{1-x^{2}}dx-\frac{2kl}{(n-1)\hbar}\biggl(\frac{El^2}{\Pi^{2/(n-3)}}\biggl)^{\alpha}\int^{x_0}_0 \frac{x^{2\alpha}}{\sqrt{1-x^{2}}}dx,
\end{align}  
where $x_0$ is defined by 
\begin{align}
x_0:=\sqrt{\frac{\Pi^{2/(n-3)}}{El^2}}y_0=\sqrt{1-d\biggl(\frac{El^2}{\Pi^{2/(n-3)}}\biggl)^{\beta-1}}.
\end{align}  
Since $x_0\to 1$ holds for large $E$, ${\bar \eta}$ is evaluated in that limit as
\begin{align}
{\bar \eta}\simeq &\frac{4El}{(n-1)\hbar}\int^{1}_0 \sqrt{1-x^{2}}dx-\frac{2kl}{(n-1)\hbar}\biggl(\frac{El^2}{\Pi^{2/(n-3)}}\biggl)^{\alpha}\int^{1}_0 \frac{x^{2\alpha}}{\sqrt{1-x^{2}}}dx \nonumber \\
=&\frac{4El}{(n-1)\hbar}\frac{\pi}{4}-\frac{kl}{(n-1)\hbar}\biggl(\frac{El^2}{\Pi^{2/(n-3)}}\biggl)^{\alpha}B\biggl(\alpha+\frac12,\frac12\biggl) \nonumber \\
\simeq &\frac{\pi lE}{(n-1)\hbar},
\end{align}  
where the equality in the second line is shown by the following expression of the Beta function:
\begin{align}
B(p,q)=2\int^{1}_0 x^{2p-1}(1-x^2)^{q-1}dx.
\end{align}  
Finally, from Eq.~(\ref{eta-x0}), we obtain 
\begin{align}
\eta(x,a)\simeq&\eta(0,a)-\sqrt{\frac{8(n-2)V_{n-2}^{(k)}E}{(n-1)^2\hbar^2\kappa_n^2}}x \nonumber \\
\simeq&{\bar \eta}-\sqrt{\frac{8(n-2)V_{n-2}^{(k)}E}{(n-1)^2\hbar^2\kappa_n^2}}x \nonumber \\
\simeq&\frac{\pi lE}{(n-1)\hbar}-\sqrt{\frac{8(n-2)V_{n-2}^{(k)}E}{(n-1)^2\hbar^2\kappa_n^2}}x \label{WKBsolution1:k=1}
\end{align}  
near $x=0$.

Comparing the above the WKB solution (\ref{WKBsolution1:k=1}) with the solution (\ref{sol:x=0}) with the Dirichlet boundary condition $B=0$ at $x=0$, we obtain the spectrum for large $E$ as
\begin{align}
&\frac{\pi lE}{(n-1)\hbar}+\frac{\pi}{4}=\pi N, \nonumber \\
\to &  E \simeq \frac{(n-1)\hbar}{l}N, 
\end{align}  
where $N$ is a positive integer.
This result is the same even when one chooses the Neumann boundary condition $A=0$ at $x=0$ and coincides with Eq.~(\ref{EN:toroidal}) in the toroidal case.

%======================================%
%<<<<<<<<<<<< SECTION I  >>>>>>>>>>>>>>%
%======================================%
\section{Summary and future prospects}
\label{sec:summary}
Adopting the throat quantization pioneered by Louko and M\"akel\"a~\cite{LM96}, we have derived the mass and area spectra for the Schwarzschild-Tangherlini black hole and its AdS generalization in arbitrary dimensions. Exact results were obtained for three special cases while the WKB approximation was used for the remainder.
For all asymptotically flat black holes, the semi-classical area/entropy spectrum turns out to be equally spaced and independent of the extensioin parameter, in agreement with ~\cite{LM96}. This is also true for the exact spectra that we derived. By contrast, in the asymptotically AdS case, we found the mass spectrum to be equally spaced.

What we calculate is the spectrum of the mass and corresponding Bekenstein/Hawking entropy for black holes with symmetry. A useful analogue  is the non-relativistic quantum mechanics of the zero angular momentum ($l=0$) hydrogen atom (i.e. a particle moving in a $1/r$ potential).
While the physically observed spectrum  ultimately requires justification from an underlying microscopic theory (QED for the Bohr atom and the as yet non-existent quantum gravity for black holes) the $l=0$ results are valid within their realm of applicability.
For example, we expect the spectra to be valid at least in the case of black holes whose mass is significantly above the Planck scale. The  mass/area spectrum is therefore potentially important in the context of black hole thermodynamics. 
As explained eloquently in the paper by Louko and M\"akel\"a, the semi-classical spectrum determines the spectrum of radiation that the black hole can emit via Hawking radiation.  A discrete mass/area spectrum of the type that we derive leads, in principle, to observable differences from the usual black body spectrum.

In addition, our calculations provide a concrete and rigorous mechanism for resolving the central singularity, again in complete analogy with the $l=0$ hydrogen atom. 
Although the resulting Hamiltonian in both cases is not essentially self-adjoint, it has a one parameter family of self-adjoint extensions~\cite{Fewster1993}.
The extension parameter, which of course does not affect the semi-classical spectrum, is a reflection of our ignorance of the underlying theory.
It may be fixed either by comparing with experimental results, when available, or more fundamentally with spectra obtained from the underlying microscopic theory.
Such an underlying theory does not exist for black holes in general, but the BTZ black hole is an exception. The quantum mechanics of the BTZ black hole has been argued (see \cite{carlip} for example) to be characterized by  a Virasoro algebra at infinity that characterizes the conformal field theory dual to the BTZ black hole in the AdS/CFT correspondence. Loosely speaking, the non-rotating BTZ black hole mass corresponds to the sum of the $L_0$ and $\overline{L}_0$ generators of this Virasoro algebra.  The spectra we derive in the case of Dirichlet and Neumann boundary conditions coincide with the one obtained for the Virasoro generators of the dual CFT. It is  encouraging that our results agree with those of the microscopic theory for a suitable choice of self-extension parameter. The fact that Dirichlet turns out to be a correct choice of boundary conditions is not surprising. It has been shown that whenever a singular potential is obtained as a limit of a (more fundamental) non-singular potential, Dirichlet boundary conditions are generic: the parameters in the non-singular potential must be fine-tuned as the limit is taken in order to obtain other boundary conditions (see \cite{walton10}, for example). We believe that the correspondence between the BTZ mass spectrum that we derived and its CFT analogue provides strong validation of our methods and suggests that both methods are correct for this simple black hole within their respective realms of validity.

In the case of higher dimensional AdS black holes the situation is less clear. The operator on the CFT side corresponding to the black hole Hamiltonian is the dilatation operator~\footnote{We are grateful to Andrew Frey for conversations in this regard.}.
The energy eigenvalues and states we have derived should therefore have analogues as eigenstates of the dilatation operator. It is therefore an interesting and highly nontrivial question whether our exact mass spectrum of the AdS black hole in higher dimensions can also be obtained using AdS/CFT duality, in analogy with what happens for the BTZ black hole. Such calculations are beyond the scope of the present work.

There are some cases which the present analysis does not cover.
One is the extremal black hole, for which the throat quantization method cannot be adopted because of the absence of a trapped region in the spacetime.
Another is the de~Sitter black hole.
In the asymptotically de~Sitter case, the absence of a timelike Killing vector at infinity prevents us from defining the conserved global mass and hence {\it asymptotically de~Sitter spacetime}.
This is the technical (and even conceptual) difficulty to show the validity of the Kucha\v{r}'s action (\ref{intro}).

We note that the generalization of the present work to the rotating Kerr-Myers-Perry black hole is problematic:
Our method is fully based on the Kucha\v{r} reduction from the vacuum action to the action with a finite number of degrees of freedom which in turn can be described by standard quantum mechanics. 
This reduction can be achieved in the case of symmetric black holes considered in the present paper as a direct consequence of the Birkhoff's theorem.
Therefore, a simple generalization to the axisymmetric black hole is not possible.

By contrast, a promising direction is to generalize our results to higher-curvature Lovelock gravity~\cite{lovelock}, which is the most natural generalization of general relativity as a quasi-linear second-derivative theory.
In throat quantization, the starting point is the simple reduced action (\ref{reduced-action}) in terms of the Misner-Sharp mass and its conjugate as canonical variables.
The same form of the reduced action has recently been derived for generic Lovelock gravity~\cite{kmt2013} in terms of the generalized Misner-Sharp mass~\cite{mwr2011}.
This result could serve as the starting point for a derivation of the energy spectrum of black holes in Lovelock gravity, for which the Bekenstein-Hawking entropy is not proportional to the area of the horizon~\cite{whitt1988}.  The results could shed light on generic properties of canonical quantum gravity.

\subsection*{Acknowledgments}
%{\bf Acknowledgments}
The authors thank Jorge Zanelli and Francisco Correa for discussions and comments. 
H.~M. thanks the Institute of Mathematics and Physics, Talca University, for its
hospitality while part of this work was carried out. G.~K. is grateful to Jorma Louko and Andrew Frey for invaluable  discussions. He also thanks CECs in Chile and  Rikkyo University in Tokyo for their kind hospitality during various stages of this work.
This research was supported in part by the Natural Sciences and Engineering Research Council of Canada and by the JSPS Grant-in-Aid for Scientific Research (A) (22244030). This work has been partially funded by the Fondecyt grants 1100328, 1100755 and by the Conicyt grant "Southern Theoretical Physics Laboratory" ACT-91. 
Centro de Estudios Cient\'{\i}ficos (CECs) is funded by the Chilean Government through the Centers of Excellence Base Financing Program of Conicyt.

\appendix

%======================================%
%<<<<<<<<<<<< SECTION I  >>>>>>>>>>>>>>%
%======================================%
\section{Natural factor ordering}
In canonical quantization, one must confront the operator-ordering problem. In the above, we follow Ref.~\cite{zanelli1986} and argue for a natural factor ordering that is obtained using geometrical considerations from the kinetic term in the Lagrangian.
Let us consider, as a simple example, a single free particle in two dimensions, of which Lagrangian and Hamiltonian in the Cartesian coordinates are
\begin{align}
L=&\frac12 m({\dot x}^2+{\dot y}^2),\\
{H} =&\frac{1}{2m}(p_x^2+p_y^2),
\end{align}  
where $p_x:=\partial L/\partial {\dot x}$ and $p_y:=\partial L/\partial {\dot y}$ are momentum conjugates.
The canonical quantization is performed by replacing as $p_x\to -i\hbar \partial/\partial x$ and $p_y\to -i\hbar \partial/\partial y$ and the resulting Schr\"odinger equation is 
\begin{align}
-\frac{\hbar^2}{2m}\biggl(\frac{\partial^2}{\partial x^2}+\frac{\partial^2}{\partial y^2}\biggl)\Psi=i\hbar \frac{\partial}{\partial t}\Psi. \label{Sch-car}
\end{align}  
Let us consider the same system in the polar coordinates:
\begin{align}
x=r\cos\theta,\quad y=r\sin\theta, \label{polar}
\end{align}  
which is a canonical transformation.
In these coordinates, the Lagrangian and the Hamiltonians are
\begin{align}
L=&\frac12 m({\dot r}^2+r^2{\dot \theta}^2),\\
{H} =&\frac{1}{2m}(p_r^2+r^{-2}p_\theta^2), \label{hamil-polar}
\end{align}  
where $p_r:=\partial L/\partial {\dot r}=m{\dot r}$ and $p_\theta:=\partial L/\partial {\dot \theta}=mr^2{\dot \theta}$ are momentum conjugates.
The canonical quantization is performed by replacing as $p_r\to -i\hbar \partial/\partial r$ and $p_\theta\to -i\hbar \partial/\partial \theta$.
Here the resulting Schr\"odinger equation must be equivalent to Eq.~(\ref{Sch-car}) and hence we have to consider a proper operator-ordering for the Hamiltonian (\ref{hamil-polar}).
For this purpose, we consider general coordinates $x^i$ in the two-dimensional flat space:
\begin{align}
ds_2^2=g_{ij}(x)dx^idx^j=dx^2+dy^2=dr^2+r^2d\theta^2.
\end{align}  
Since the left-hand side of Eq.~(\ref{Sch-car}) is the Laplacian operator in the two-dimensional flat space, Eq.~(\ref{Sch-car}) can be written as
\begin{align}
&-\frac{\hbar^2}{2m}g^{ij}\partial_i\partial_j\Psi=i\hbar \frac{\partial}{\partial t}\Psi, \label{Sch-car2}\\
\to &-\frac{\hbar^2}{2m}\frac{1}{\sqrt{g}}\frac{\partial}{\partial x^i}\biggl(\sqrt{g}g^{ij}\frac{\partial}{\partial x^j}\biggl)\Psi=i\hbar \frac{\partial}{\partial t}\Psi.
\end{align}  
In the polar coordinates, it becomes
\begin{align}
&-\frac{\hbar^2}{2m}\frac{1}{r}\biggl(\frac{\partial}{\partial r}\biggl(rg^{rr}\frac{\partial}{\partial r}\biggl)+\frac{\partial}{\partial \theta}\biggl(rg^{\theta\theta}\frac{\partial}{\partial \theta}\biggl)\biggl)\Psi=i\hbar \frac{\partial}{\partial t}\Psi, \nonumber \\
\to &-\frac{\hbar^2}{2m}\biggl(\frac{1}{r}\frac{\partial}{\partial r}\biggl(r\frac{\partial}{\partial r}\biggl)+r^{-2}\frac{\partial^2}{\partial \theta^2}\biggl)\Psi=i\hbar \frac{\partial}{\partial t}\Psi.
\end{align}  
The corresponding ordering for the Hamiltonian (\ref{hamil-polar}) is 
\begin{align}
{H} =&\frac{1}{2m}(r^{-1}p_r r p_r+r^{-2}p_\theta^2).
\end{align}  
This is what we call the {\it natural ordering}.
Under this ordering, a natural measure can be introduced:
\begin{align}
\langle \Phi,\Psi\rangle:=\int\Phi^\ast\Psi \sqrt{g}d^2x.
\end{align}  

The metric $g_{ij}$ can be read off from the form of the Hamiltonian (\ref{hamil-polar}) because it has the form of
\begin{align}
{H} =&\frac{1}{2m}g^{ij}p_ip_j, \label{hamil-polar2}
\end{align}  
from which we obtain the non-zero components of $g^{ij}$:
\begin{align}
g^{rr}=1,\quad g^{\theta\theta}=r^{-2}
\end{align}  
and hence we obtain 
\begin{align}
g_{rr}=1,\quad g_{\theta\theta}=r^{2}.
\end{align}

%======================================%
%<<<<<<<<<<<< SECTION I  >>>>>>>>>>>>>>%
%======================================%
\section{Proof that energy is bounded below}
\label{section:energy bound}
Let us consider the time-independent Schr\"odinger equation $\hat
H\psi(x)=E\psi(x)$ for the following Hamiltonian operator:
\begin{align}
\hat H=-\frac{\hbar^2}{2m}\frac{d^2}{dx^2}+ V(x),
\end{align}
where $m>0$.
The self-adjointness boundary condition requires
\begin{align}
\psi (0) + L \psi'(0) = 0,
\end{align}
where $L$ is a constant and a prime denotes derivative with respect to $x$.
This is the same as Louko and M\"akel\"a~\cite{LM96} for $L \equiv
\tan\theta$.
We assume that the potential $V(x)$ is continuous and bounded below as
$V(x) \ge V_{\rm min}$ for $0\le x<\infty$.
We will show, by contradiction, that the eigenvalue $E$ for the above
Hamiltonian operator is also bounded below.

First we consider the case of $-\infty<L \le 0$.
Suppose $E<V_{\rm min}$ and write the Schr\"odinger equation as
\begin{align}
\frac{\hbar^2}{2m}\frac{d^2\psi}{dx^2}= \left(V(x)-E\right)\psi.
\label{eigen-key1}
\end{align}
The solutions around $x=0$ that have the right boundary condition take
the form of
\begin{align}
\psi(x)\simeq &A\biggl((1-\lambda L)e^{\lambda x}-(1+\lambda
L)e^{-\lambda x}\biggl) \label{asymp-x}
\end{align}
which gives
\begin{align}
\psi(x)\psi'(x)\simeq &A^2\lambda \biggl((1-\lambda L)^2e^{2\lambda
x}-(1+\lambda L)^2e^{-2\lambda x}\biggl),
\end{align}
where $A$ is a constant and $\lambda:=\sqrt{2m(V(0)-E)/\hbar^2}$.
Note that $\lambda$ is real and positive because of $E<V_{\rm min}\le V(0)$.
Using the phase ambiguity of $\psi(x)\to \psi(x)e^{i\varphi}$ with a
real parameter $\varphi$, we can set $A$ real without loss of generality.

It is seen that $\psi(x)\psi'(x)>0$ is satisfied for $-\infty<L \le 0$
and then the signs of $\psi(0)$ and $\psi'(0)$ are the same.
Since $V(x)-E$ is non-negative, the sign of $\psi''(0)$ is the same as
those of $\psi(0)$ and
$\psi'(0)$.
Now as one moves away from the origin, $|\psi(x)|$ will continue to
increase and therefore cannot be normalizable.
Hence, by contradiction, $E\ge V_{\rm min}$ is satisfied if $-\infty<L
\le 0$.
This argument does work also for $L=\pm \infty$ corresponding to the Neumann
boundary condition $\psi'(0)=0$ because then we have
\begin{align}
\psi(x)\simeq &A(e^{\lambda x}+e^{-\lambda x}) \label{asymp-x2}
\end{align}
which gives
\begin{align}
\psi(x)\psi'(x)\simeq 4A^2\lambda^2 x>0.
\end{align}

In the case of $0<L <\infty$, on the other hand, we show that $E$ is
just bounded below.
Suppose that $E$ is unbounded.
If $E<V(0)$, the asymptotic solution (\ref{asymp-x}) is valid with real
and positive $\lambda$ and $\psi(x)\psi'(x)>0$ is satisfied for
$x>x_{\rm min}$, where
\begin{align}
x_{\rm min}:=\frac{1}{2\lambda}\ln \biggl|\frac{1+\lambda L}{1-\lambda
L}\biggl|(>0).
\end{align}
For a sufficiently large value of $|E|$, $V(x)-E$ is non-negative and
there is a region near $x=0$ where $\psi(x)\psi'(x)>0$ holds because of
$\lim_{\lambda\to \infty}x_{\rm min}=1/(\lambda^2 L)\to 0$.
Then, by the same argument for $-\infty<L \le 0$ in such a region, it is
concluded that $\psi(x)$ cannot be normalizable and therefore $E$ is
bounded below.
The result presented here is consistent with the exact results for the
simple harmonic oscillator on the half line given in the following
Appendix~\ref{appendix:half-line}.

%======================================%
%<<<<<<<<<<<< SECTION I  >>>>>>>>>>>>>>%
%======================================%
\section{Quantum harmonic oscillator on the half line}
\label{appendix:half-line}
In this appendix, we review quantization of a one-dimensional harmonic oscillator on the half line\footnote{Taken from an old draft of the paper~\cite{kl2012} with thanks to Jorma Louko.}.
Let us consider the Hamiltonian for a classical single harmonic oscillator:
\be
H(x,p)= \frac{1}{2m}p^2+\frac12m\omega^2x^2,
\ee
where $m$ and $\omega$ are positive constants.
The resulting Schr\"odinger equation with standard measure acting on normalizable states $\Psi(t,x)=e^{-i{\bar E}t/\hbar}\psi(x)$ is
\be
\biggl(-\frac{\hbar^2}{2m}\frac{d^2}{dx^2} + \frac12 m\omega^2x^2\biggl) \psi = {\bar E} \psi.
\ee
By the scaling transformation $x=(\hbar/m\omega)^{1/2}\xi$, we obtain
\be
\biggl(-\frac{d^2}{d\xi^2} + \xi^2\biggl) \psi(\xi) = 2E \psi(\xi), \label{eq:se}
\ee
where $E:={\bar E}/m\omega$.

If the domain of the coordinate is $0\leq \xi<\infty$, then we need to specify a boundary condition at the origin $\xi=0$ that preserves probability, rules out imaginary eigenvalues and still leaves a non-trivial solution space. The family of suitable boundary conditions is:
\be
\psi(0) \cos\chi+ \psi'(0) \sin\chi = 0
\label{eq:bc origin1}
\ee
with $\chi\in [0,\pi)$, where a prime denotes the derivative with respect to $\xi$. 
This can be related to the boundary condition derived directly from making sure the probability flow throw the origin is zero, namely that:
\be
\psi(0) + L \psi'(0) = 0
\ee
for any real $L$. Clearly $L \equiv  \tan\chi$ spans the real numbers for the given range of $\chi$. However, the first firm takes into account that the parameter essentially specifies a phase change on reflection and hence is periodic with respect to its effect on the spectrum.

The solutions to Eq.~(\ref{eq:se}) are in general written in terms of the parabolic cylinder functions 
$U(a,\xi)$ found in Section 12.2 in~\cite{NIST}.  
Equation (\ref{eq:se}) corresponds to 12.2.2 in~\cite{NIST} with the identifications $a \equiv -E$ and $z \equiv \sqrt{2}\xi$. 
The boundary conditions (\ref{eq:bc origin1}) will determine the energy spectrum, making use of 12.2.6 and 12.2.7 in~\cite{NIST}:
\bea
U(a,0) &=& \frac{\sqrt{\pi}}{2^{a/2+1/4}\Gamma(3/4+a/2)},\\
U'(a,0) &=& -\frac{\sqrt{\pi}}{2^{a/2-1/4}\Gamma(1/4+a/2)}.
\eea
The boundary conditions (\ref{eq:bc origin1}) is now found to be
\bea
&&\frac{\cos\chi}{\Gamma(3/4-E/2)}-\frac{\sqrt{2}\sin\chi}{\Gamma(1/4-E/2)}=0,\\
\to &&\tan\chi=\frac{\Gamma(1/4-E/2)}{\sqrt{2}\Gamma(3/4-E/2)}=:g(E). \label{def-g}
\eea
%\be
%g(1/4-E/2):=\sqrt{2}\tan\chi{\Gamma\left(\frac14-\frac{E}{2}+\frac12\right)}-{\Gamma\biggl(\frac14-\frac{E}{2}\biggl)}=0
%\label{eq:sho_robin}
%\ee
\begin{figure}[htbp]
\begin{center}
%\rotatebox{-90}{
\includegraphics[width=0.7\linewidth]{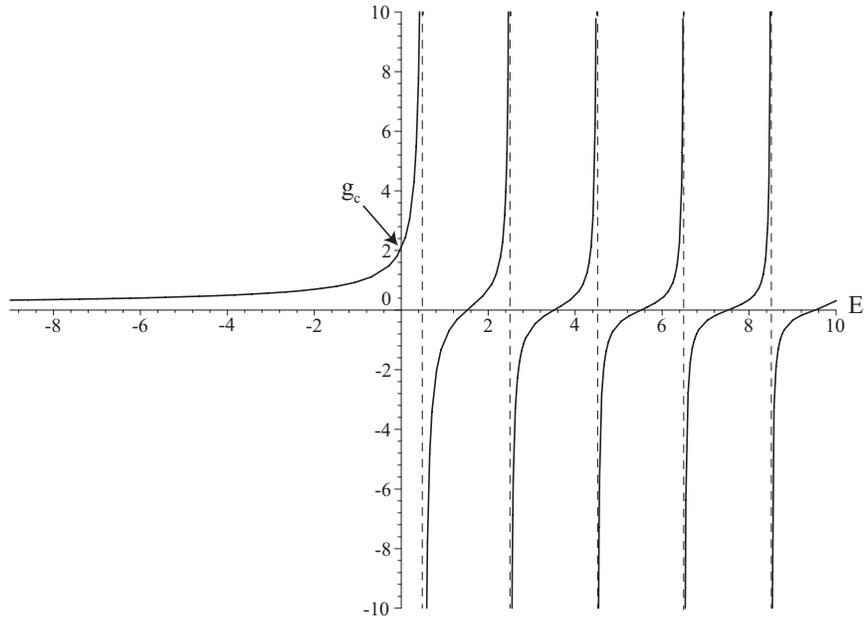}
%\subfigure[]{\includegraphics[width=0.7\linewidth]{Roberts-lambda1.eps}}
%\subfigure[]{\includegraphics[width=0.7\linewidth]{Roberts-lambda2.eps}}
%}
\caption{\label{fig:eh} Plot of $g(E)$ from Eq.~(\ref{def-g}) as a function of $E$, where $g_{\rm c}:=g(0)=\Gamma(1/4)/\sqrt{2}\Gamma(3/4)$.
The system allows the negative eigenvalue of $E$ if we choose the boundary condition with $\chi$ satisfying $0<\tan\chi<g_{\rm c}$.}
\end{center}
\end{figure}

$\chi=\pi/2$ corresponds to symmetric (Neumann) boundary conditions, and the condition that determines the energy eigenvalues is:
\bea
\Gamma\biggl(\frac14-\frac{E}{2}\biggl)=\infty\quad \rightarrow &&\frac14 -\frac{E}{2} = -N \quad (N=0,1,2,\cdots) \nonumber\\
\rightarrow &&E = 2N + \frac{1}{2}=:E_{\rm s},
\eea
which are indeed the eigenvalues of the symmetric states of the single harmonic oscillator.
Similarly $\chi =0$ gives Dirichlet boundary conditions and the anti-symmetric eigenvalues, namely:
\be
E= (2N+1) + \frac{1}{2}=:E_{\rm a}.
\ee
In these cases, the energy is equally spaced.
However, it is not the case in general.

The energy spectrum for more general $\chi$ can be read off from Fig.~\ref{fig:eh}.
Cearly $E$ is not equally spaced for general $\chi$.
Moreover, the ground state is negative for the boundary conditions with $\chi$ satisfying $0<\tan\chi<g_{\rm c}$.

It is important to keep in mind that each value of $\chi$ corresponds to a different, inequivalent quantization and  no transitions between different boundary conditions are allowed.

%%%%%%%%%%%%%%%%%%%%%%%%%%%%%%%%%%%%%%%%%%%%%%%%%%%%%%%%%%%%%%%%%%%%%%%%
\end{document}